\documentclass[12pt]{iopart}
\usepackage{graphicx}
\usepackage{float}
\usepackage[geometry]{ifsym}
\usepackage{color}
\usepackage{hhline}
\usepackage{array}
\usepackage{setspace}

\begin{document}

\title{Experimental study of Counter-Rotating Vortex Pair Trajectories induced by a Round Jet in Cross-Flow at Low Velocity Ratios}
\author{T. Cambonie, N. Gautier, J.-L. Aider }
\address{PMMH, 10, rue Vauquelin 75006 Paris, France}

\begin{abstract}
Circular flush Jets In Cross-Flow were experimentally studied in a water tunnel using Volumetric Particle Tracking Velocimetry, for a range of jet to cross-flow velocity ratios, $r$, from $0.5$ to $3$,  jet exit diameters $d$ from $0.8$ $cm$ to $1$ $cm$ and cross-flow boundary layer thickness $\delta$ from $1$ to $2.5$ $cm$. The analysis of the 3D mean velocity fields allows for the definition, computation and study of Counter-rotating Vortex Pair trajectories. The influences of $r$, $d$  and $\delta$ were investigated. A new scaling based on momentum ratio $r_m$ taking into account jet and cross-flow momentum distributions is introduced based on the analysis of  jet trajectories published in the literature. Using a rigorous scaling quality factor $Q$ to quantify how well a given scaling successfully collapses trajectories, we show that the proposed scaling also improves the collapse of CVP trajectories,  leading to a final scaling law for these trajectories.
\end{abstract}


\section{Introduction}
\label{intro}
Jets In Cross-Flows (JICF) are complex three-dimensional flows which can be found in many engineering applications such as film cooling of turbines and combustors or the control of separated flows over airfoils and ground vehicles (\cite{Margarson1993}, \cite{Stanislas2006}, \cite{Joseph2011}). The control and understanding of JICF's is of great industrial interest. Its complexity  also makes it a great challenge for academic research. Thus, it has been the subject of many experimental, numerical and theoretical studies over the past fifty years which are well summarized in the recent review by \cite{AnnR2010}. 

When studying a JICF, many parameters can be considered, such as the Reynolds numbers of both jet and cross-flow, the diameter of the jet or the velocity ratio. The latter is considered as the key parameter and is defined as $r=\sqrt{\rho_j \overline{V_j}^2/\rho_{\infty} U_{\infty}^2}$ where $\rho_j$,$V_j$ are the jet density and  mean exit velocity and $\rho_{\infty}$,  	 $U_{\infty}$ are the free stream density and velocity. When jet and free stream fluid densities are equal, the momentum ratio becomes $r=\overline{V_j}/U_{\infty}$.

\begin{figure}
\centering
\centerline{\includegraphics[width=1\textwidth]{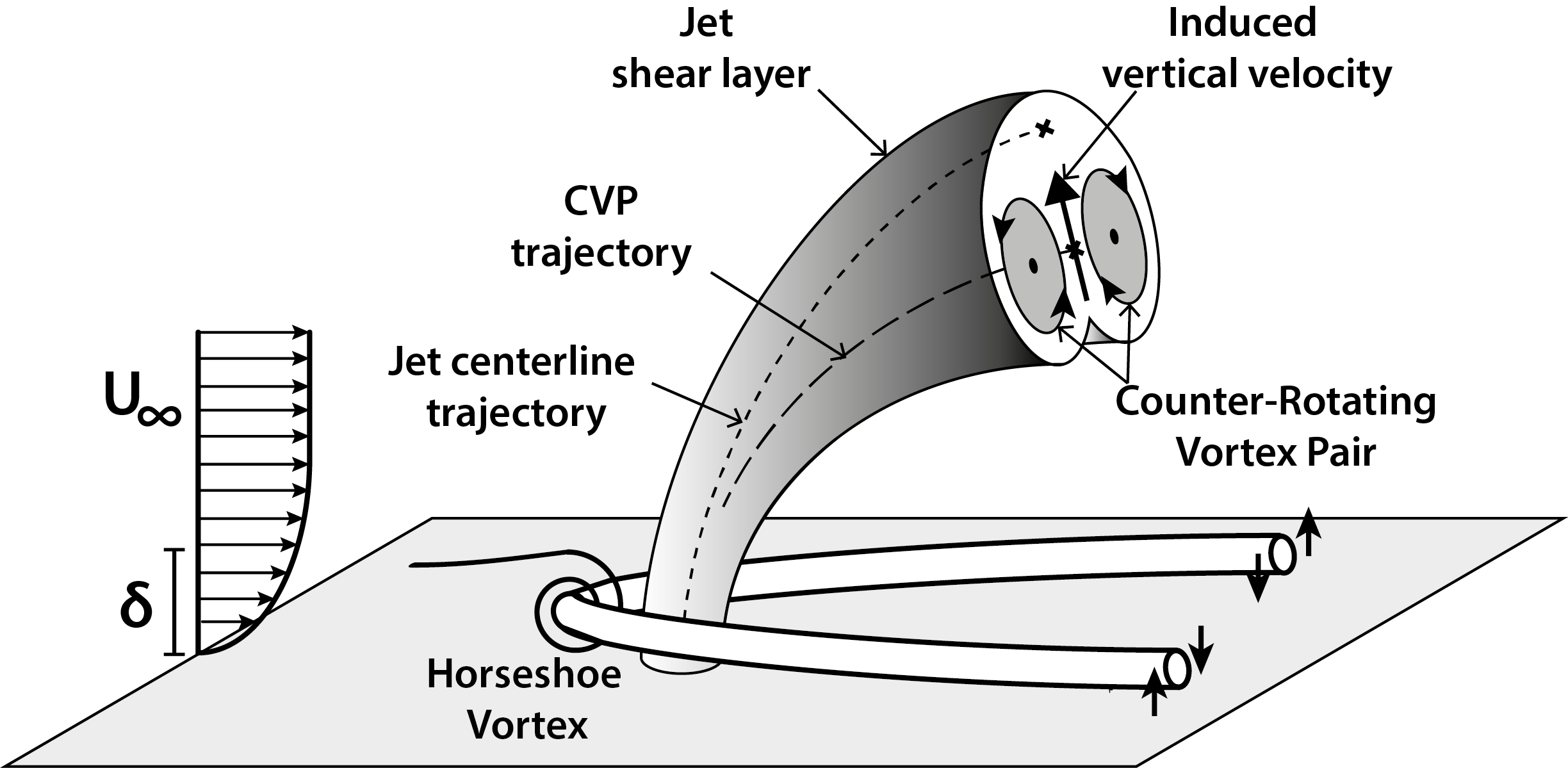}}
\caption{ Sketch of jet in cross-flow: the CVP and the Horseshoe vortex are the main swirling structures observed in the mean velocity field.}
\label{fig:SketchCVP}
\end{figure}

The main feature of the mean flow observed in previous studies is the counter-rotating vortex pair (CVP), sketched on Fig. \ref{fig:SketchCVP}. CVP are, to our knowledge, always present in time-averaged velocity fields. Moreover, the CVP is the only structure remaining far from the injection site, sometimes persisting as far as a thousand jet diameters as shown by \cite{Keffer1963}.  The CVP has been investigated in detail by \cite{Chassaing1974}, \cite{Blanchard1999}, \cite{Cortelezzi2001} and \cite{Marzouk2007}. Characterization of its location through the study of its trajectory is therefore of great interest.

We consider low velocity ratios ($r<3$). Most previous studies focused on higher velocity ratios ($r>2-3$). Low velocity ratios JICF's were investigated by \cite{Camussi2002} and \cite{Gopalan2004}. A significant difference between high and low velocity ratios is the interaction with the boundary layer: at low $r$ the jet interacts with the boundary layer leading to a  profound modification of the flow structure. Transition between globally unstable and convectively unstable flow has been shown to exist at $r=3$ by \cite{Megerian2007}.  A transition at very low velocity ratios ($r=0.3$) has been observed by \cite{Cambonie2012}. It is a transition from a \textit{blown} jet topology to a \textit{classical} jet topology. These transitions could impact the CVP. Our range of velocity ratios  is $0.5<r<3$, above the transition from blown jet to classical jet.
\\\\
To our knowledge there are no parametric studies focusing on CVP trajectory, although they are mentioned as vortex curves and studied by \cite{Weston1974} and \cite{Karagozian1986}. The objective of this paper is to define the CVP trajectories in such a way that it can be computed for any velocity ratio and to propose a scaling for these trajectories which takes  into account jet and boundary layer momentum distributions,  cross-flow boundary layer thickness and jet diameter for low velocity ratios.

\section{Experimental setup}
\label{sec:1}
\subsection{Water tunnel, jet supply system and geometries}
\label{sec:2}
\begin{figure}
\centering
\centerline{\includegraphics[width=0.95\textwidth]{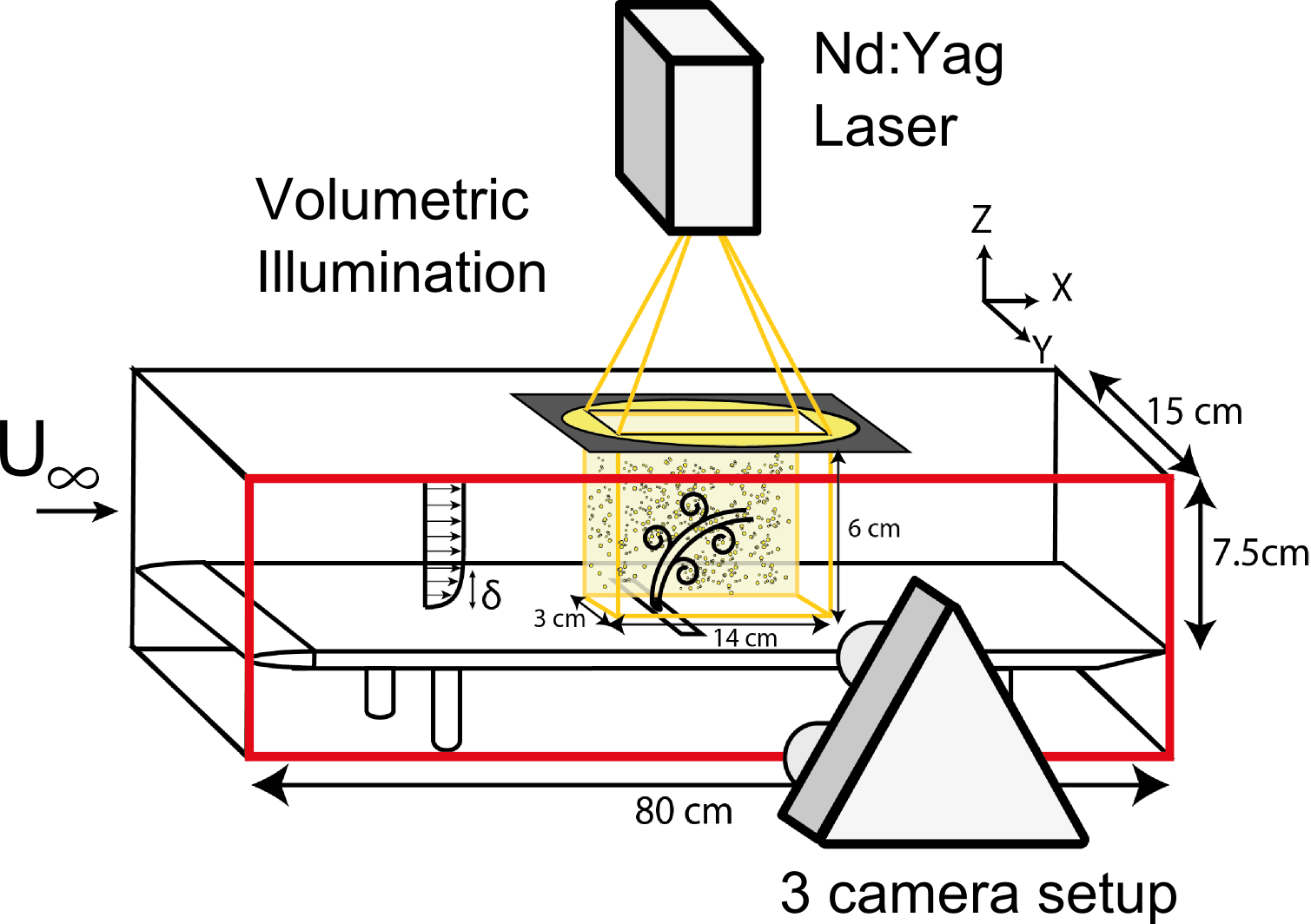}}
\caption{Definition of the experimental test section. The flow goes from left to right and develops over a raised plate with NACA leading edge. The measurement volume is lit through the upper plate. The three cameras of the V3V system are tracking particles through the side-wall of the channel. The jet nozzle is located $42 cm$ downstream of the leading edge.}
\label{fig:plate}
\end{figure}

Experiments were conducted in a hydrodynamic channel in which the flow is driven by gravity. The walls are made of Altuglas for easy optical access from any direction. Upstream of the test section the flow is stabilized by divergent and convergent sections separated by honeycombs. The test section is $80$  $cm$ long with a rectangular cross section $15$ $cm$ wide and $20$ $cm$ high as described in Fig. \ref{fig:plate}.\\

 The mean free stream velocity $U_{\infty}$ ranges between $0.9$ to $8.37$ $cm.s^{-1}$ corresponding to $Re_{\infty}=\frac{U_{\infty}d}{\nu}$ ranging between 220 and 660.   The quality of the main stream can be quantified in terms of flow uniformity and turbulence intensity. 
The spatial $\sigma_s$ and temporal $\sigma_t$ standard deviations are computed using a sample of 600 velocity fields. The values are, for the highest free stream velocity featured in our data, $\sigma_s=0.038$ $cm.s^{-1}$ and $\sigma_t=0.059$ $cm.s^{-1}$ which corresponds to turbulence levels $\frac{\sigma_s}{U_{\infty}}=0.15$ $\%$ and  $\frac{\sigma_t}{U_{\infty}}=0.23$ $\%$, respectively.
\\
A custom made plate with a specific leading-edge profile is used to start the cross-flow boundary layer.   The boundary layer over the plate is laminar and stationary according to $Re_x=\frac{U_{\infty}x}{\nu}<2100$, where x is the distance to the leading edge of the plate, for the highest free stream velocity case, which is considerably less than the critical value for this profile. The boundary layer characteristics were investigated using 600 instantaneous 3D velocity fields without a jet present for all cross-flow velocities. The average field allows us to compute the boundary layer velocity profiles. The boundary layer thickness $\delta$ varies from $2.5$ $cm$ to $1$  $cm$ for increasing cross-flow velocity.
\\
  These unperturbed fields were used to compute cross flow velocity by averaging longitudinal velocity in the volume field, excluding the boundary layer.

\begin{figure}
\centering
\includegraphics[width=0.5\textwidth]{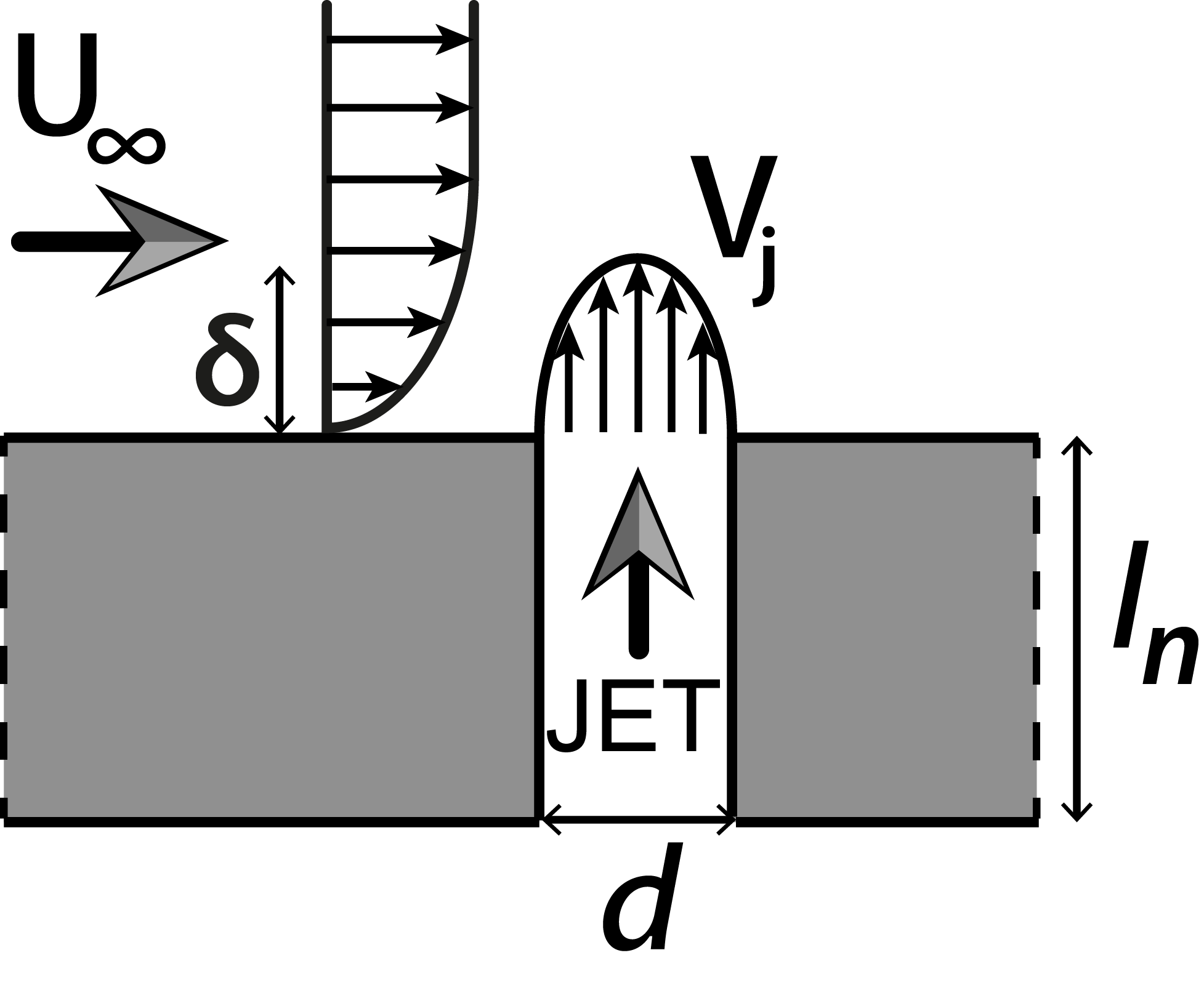}
\caption{2D sketch of the injection site with definitions of the main geometric and physical parameters.}
\label{fig:nozzle}
\end{figure}

The jet supply system was custom made. Water enters a plenum and goes through a volume of glass beads designed to homogenize the incoming flow. The flow then goes through a cylindrical nozzle which exits flush into the cross-flow. In the following, we focus on nozzles with different diameters $d$ and different injection lengths $l_n$  (Fig. \ref{fig:nozzle}, and table \ref{tab:configs}). The jet axis is normal to the flow. The mean vertical jet velocity $\overline{V_{j}}$ ranges between $1.9$ and $8$ $cm.s^{-1}$, leading to velocity ratios $r = \overline{V_{j}} / U_{\infty}$ ranging between $0.5$ to $3$. The dimensions of the jet nozzle and flow characteristics for the 22 configurations presented in this study are summarized in table \ref{tab:configs}.

\begin{table}[htb!]
\footnotesize{
\begin{tabular}{|>{\centering}m{2cm}| c c c c c c c c c c c |}
\hline
{Configuration\newline Number}&1&2&3&4&5&6&7&8&9&10&11 \\ \hline
{$d (cm)$}&0.8&0.8&0.8& 0.8 &0.8&0.8&0.8&0.8&0.8&1&1 \\
{$l_n (cm)$}&1&1&1&1 &1&1&1&1&1&1&1\\
{$U_{\infty}(cm.s^{-1})$}&5.58&4.13&2.66&1.87 &6.53&6.51&6.54&6.39&4.01 &2.57&1.70\\
{$\overline{V_j}(cm.s^{-1})$}&3.04&3.04&3.04&3.04 &3.31&5.40&7.00&7.95&6.25&3.87&2.85\\
{$\delta/d$} &1.83&1.96&2.31&2.7&1.78&1.78&1.78&1.78&1.59&1.87&2.25\\
{$Re_{\infty}$}&450&330&210&150&520&520&520&510&400&260&170\\
{$r$}&0.54&0.74&1.14&1.62 &0.51&0.83&1.07&1.24&1.56&1.51&1.67\\
{$r_m$}&0.75&1.01&1.58&2.23&0.70&1.13&1.45&1.68&2.11&2.105&2.29\\
{Markers} &-\FilledCircle & $-\times$ & $-\bullet$ & $-+$ & -\FilledTriangleDown & -\FilledTriangleUp & $- -$ & $\cdot \cdot \cdot$ & $- \cdot -$ & $-$& -\FilledDiamondshape
\\\hline\multicolumn{12}{c}{}\\\hline

{Configuration\newline Number}&12&13&14&15&16&17&18&19&20&21&22\\ \hline
{$d (cm)$}&1&1&1&1&1&1&1&1&1&1&1\\
{$l_n (cm)$}&1&1&0.5&0.5&0.5&2&2&2&3&3&3\\
{$U_{\infty}(cm.s^{-1})$}&1.23&1.07&6.58&3.20&2.06&6.55&3.24&2.09&6.55&3.25&2.17\\
{$\overline{V_j}(cm.s^{-1})$}&2.02&1.71&6.30&6.30&6.30&6.30&6.30&6.30&6.30&6.30&6.30\\
{$\delta/d$}&2.59&2.72&1.41&1.71&2.06&1.42&1.70&2.05&1.42&1.70&2.02\\
{$Re_{\infty}$}&120&110&660&320&210&650&320&210&660&330&220\\
{$r$}&1.64&1.59&0.96&1.97&3.05&0.96&1.94&3.01&0.96&1.94&2.9\\
{$r_m$}&2.27&2.20&1.29&2.65&4.10&1.32&2.66&4.12&1.32&2.67&4.00\\
{Markers} &-\FilledCircle & $-\times$ & $-\bullet$ & $-+$ & -\FilledTriangleDown & -\FilledTriangleUp & $- -$ & $\cdot \cdot \cdot$ & $- \cdot -$ & $-$& -\FilledDiamondshape\\\hline

\end{tabular}}
 \caption{The 22 configurations are defined by a set of eight parameters: jet diameter, injection length, free stream velocity, jet velocity, boundary layer thickness, Reynolds number and momentum ratios. The markers associated to each configuration, and used in the following figures, are also defined.}
\label{tab:configs}
\end{table}

\subsection{3D Particle Tracking Velocimetry measurements}
\label{sec:3}
To analyze the mean-flow characteristics of the JICF, we use volumetric particle tracking velocimetry (3DPTV). The method was pioneered by  \cite{WILLERT1992}  and further developed by  \cite{Pereira2002}. The set-up was designed and the physical parameters were chosen to optimize the quality of the instantaneous velocity fields, using the methodology of \cite{Cambonie2013}.   We used $50\mu m$ polyamide particles (PSP) for seeding, with a concentration of $5.10^{-2}$ particles per pixel.  The flow is illuminated through the upper wall and the particles are tracked using three cameras facing the side wall  (Fig. \ref{fig:plate}). The three double-frame cameras are $4$ MP with a 12 bit output. Volumetric illumination is generated using a 200 mJ pulsed YaG laser and two perpendicular cylindrical lenses. Synchronization is ensured by a TSI synchronizer. The measurement volume $(l_x,l_y,l_z)$ is $14 \times 6 \times 3$ $cm^3$. The spatial resolution is one velocity vector per millimeter for both the instantaneous and mean three-components velocity field .  This resolution might not always allow for the detection of the smallest structures in the flow, especially for higher velocity ratios. Nevertheless the jet diameter has been chosen to ensure a good spatial resolution of the main vortices ($8$ $mm<d<10$ $mm$). The characteristic width of a vortex is the an order of magnitude higher than the spatial resolution allowing us to clearly detect the CVP. The acquisition frequency is 7.5$Hz$. 1000 instantaneous velocity fields are recorded for each configuration to ensure statistical convergence of the mean velocity field.

\section{Trajectory computation}
\label{sec:5}
\subsection{  Visualization of the CVP}
To analyze the complex three-dimensional flow, we use the swirling strength criterion $\lambda_{ci}$. It was first introduced by \cite{Chong1990} who analyzed the velocity gradient tensor and proposed that the vortex core be defined as a region where $\nabla u$ has complex eigenvalues.   It was later improved and used for the identification of vortices in three-dimensional flows by \cite{Zhou1999}. This criterion allows for an effective detection of vortices even in the presence of shear. It is calculated for the entire 3D velocity fields.
\begin{figure}
\centerline{\includegraphics[width=0.95\textwidth]{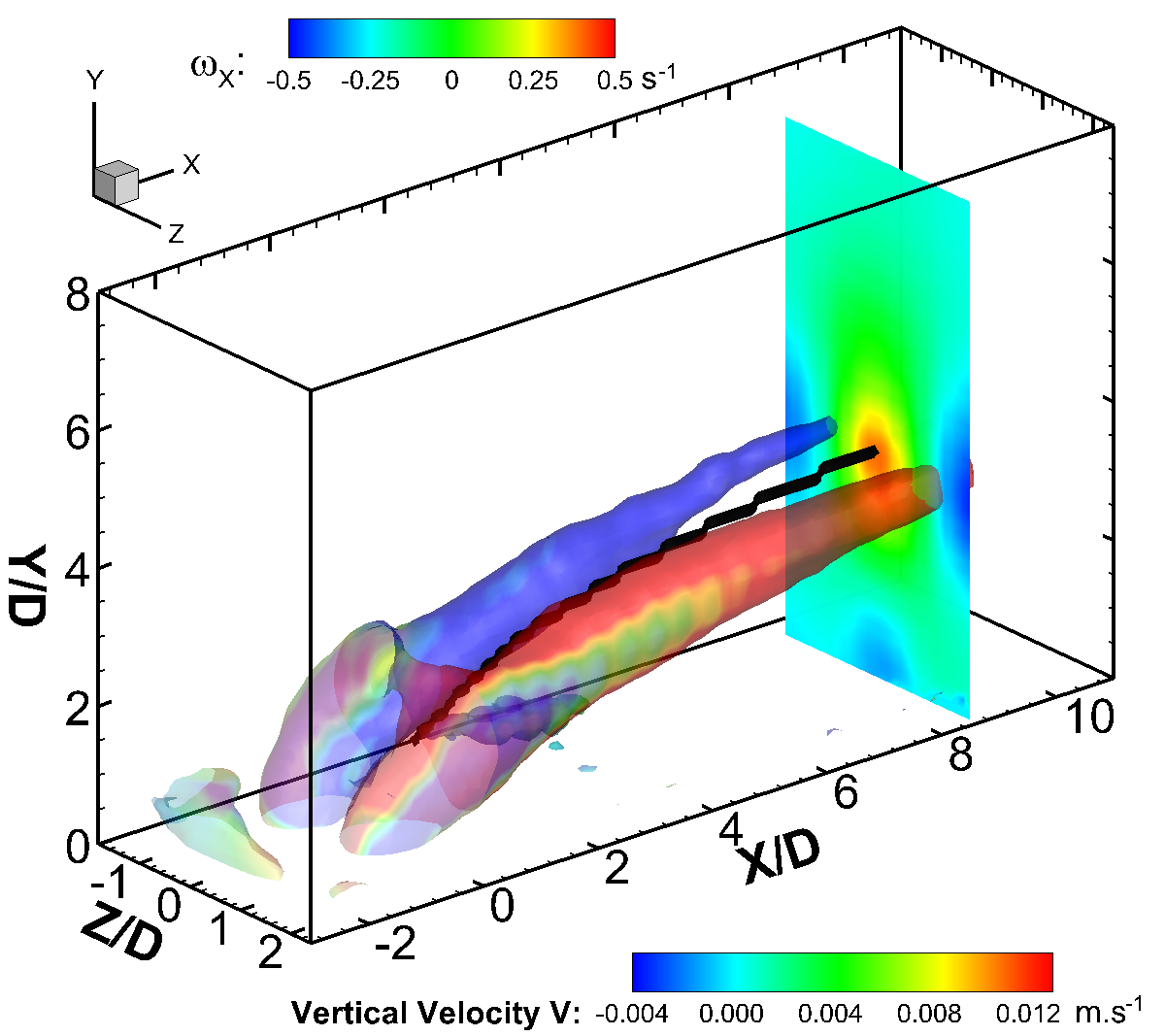}}
\caption{Mean iso-surface of $\lambda_{ci}$ colored by longitudinal vorticity for configuration 7 (velocity ratio $r=1.07$, together with a contour of vertical velocity at $X=10d$. The computed CVP trajectory is shown as a thick black line.}
\label{fig:3DTraj}
\end{figure}
Fig. \ref{fig:3DTraj} shows a typical example of the main vortical structures present in the mean velocity field using isosurfaces of $1.5\cdot \sigma(\lambda _{Ci}) $ (where $\sigma$ is the spatial standard deviation) colored by the longitudinal vorticity.  One can clearly see the two counter-rotating vortices growing downstream of the injection site. The vertical velocity field is also visualized in the $X/d=10$ cross-section showing the strong outflow region induced by the CVP.    The CVP creates a well-defined outflow region in its center. Thus a practical way of computing the CVP trajectory is to look for the locus of maximal vertical velocity.

\subsection{ Jet and CVP trajectory}
\label{sec:6}

\begin{figure}
\centering
\includegraphics[width=0.75\textwidth]{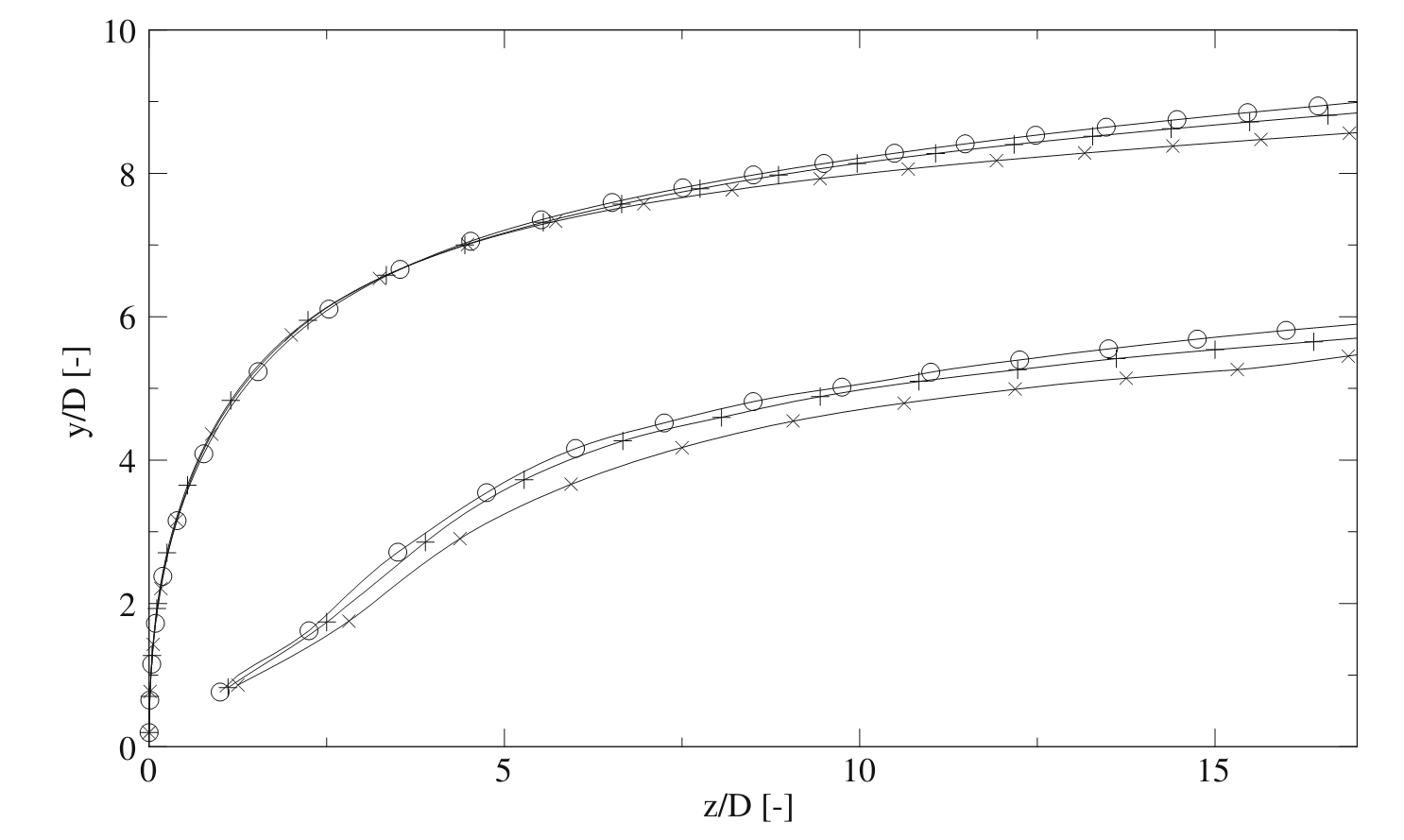}
\caption{Jet centerline trajectories (higher curves), CVP trajectories (lower curves) computed in the numerical simulations of  \cite{Salewski2007}. To the different trajectories correspond different number of cells for the numerical simulations:  3.8 million cells (-\FilledCircle), 3.2 million cells ($+$)  and  2.4 million cells ($\times$).}
\label{fig:CVPunder}
\end{figure}

It is important to stress that CVP trajectory and jet trajectory are distinct entities.  \cite{Muppidi2007}, \cite{Salewski2007}, as well as \cite{Mungal2001} show that the CVP trajectory lies under the jet trajectory. \\
  There are several ways of defining the jet trajectory: the jet centerline (for circular jets it is the streamline starting at the center of the injection nozzle), the locus of maximum velocity or the locus of maximum concentration. \cite{Yuan1998} compare these methods and show that although the computed trajectories vary, they show the same behavior.
\\ 
Fig. \ref{fig:CVPunder} (\cite{Salewski2007}) features numerical data showing the jet centerline trajectory and the location of the CVP. The CVP does not start at origin ($x=0, y=0$) and is clearly lower than the jet centerline. However both trajectories are parallel for $z/d>8$. This is because the CVP is a structure of the mean flow field, a time average of transient structures in the instantaneous flow as shown by \cite{FRIC1994}. 
  
\begin{figure}
\begin{center}
\begin{tabular}{c c}
\includegraphics[width=0.55\textwidth]{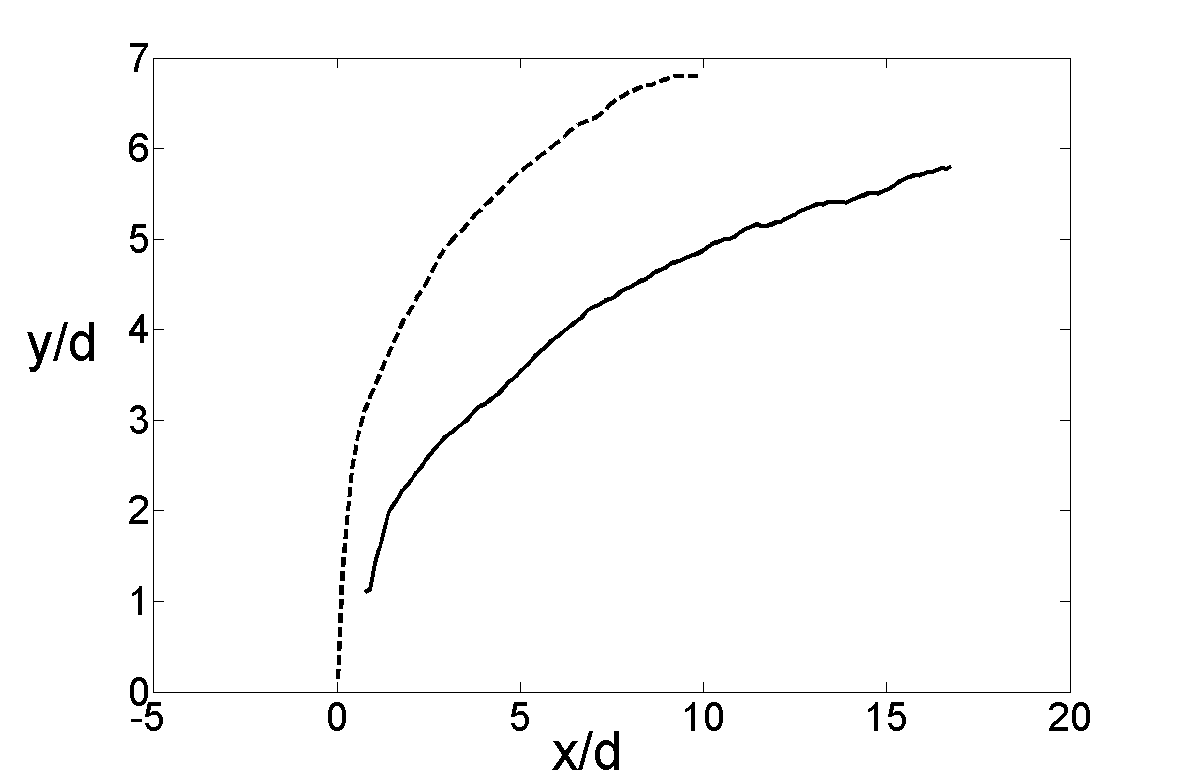} & \includegraphics[width=0.55\textwidth]{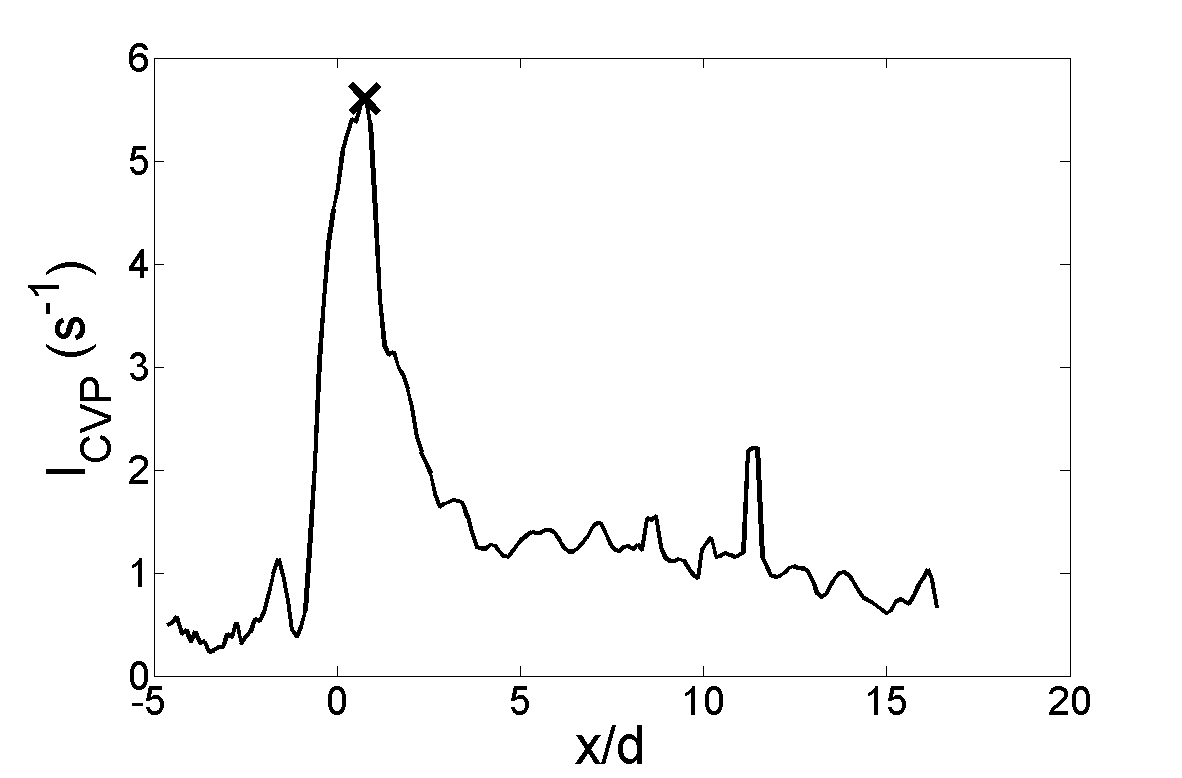}\\
a) & b)
\end{tabular}
\end{center}
\caption{  a) Jet (- -) and CVP(-) trajectories for configuration 10 ($r=1.51$). b) Swirling strength of the strongest vortex for configuration 10. Maximum is indicated by a cross.}
\label{fig:JetvsCVPtraj}
\end{figure}
\label{sec:7}

When the velocity ratio $r$ is high enough, the difference between jet and CVP trajectories can be observed in our data, as shown by Fig. \ref{fig:JetvsCVPtraj} a. To compute these trajectories we locate the two vertical velocity maxima in every cross sections. This gives the (y-z)-coordinates of the CVP and jet trajectory for the given abscissa. This computation method is straightforward, easy to implement and applicable at any velocity ratio. It allows us to distinguish vertical velocity created by the CVP and vertical velocity from the jet itself.

\subsection{  Computing CVP trajectories}
\label{sec:8}
\begin{figure}
\begin{center}
\begin{tabular}{c c}
\includegraphics[width=0.5\textwidth]{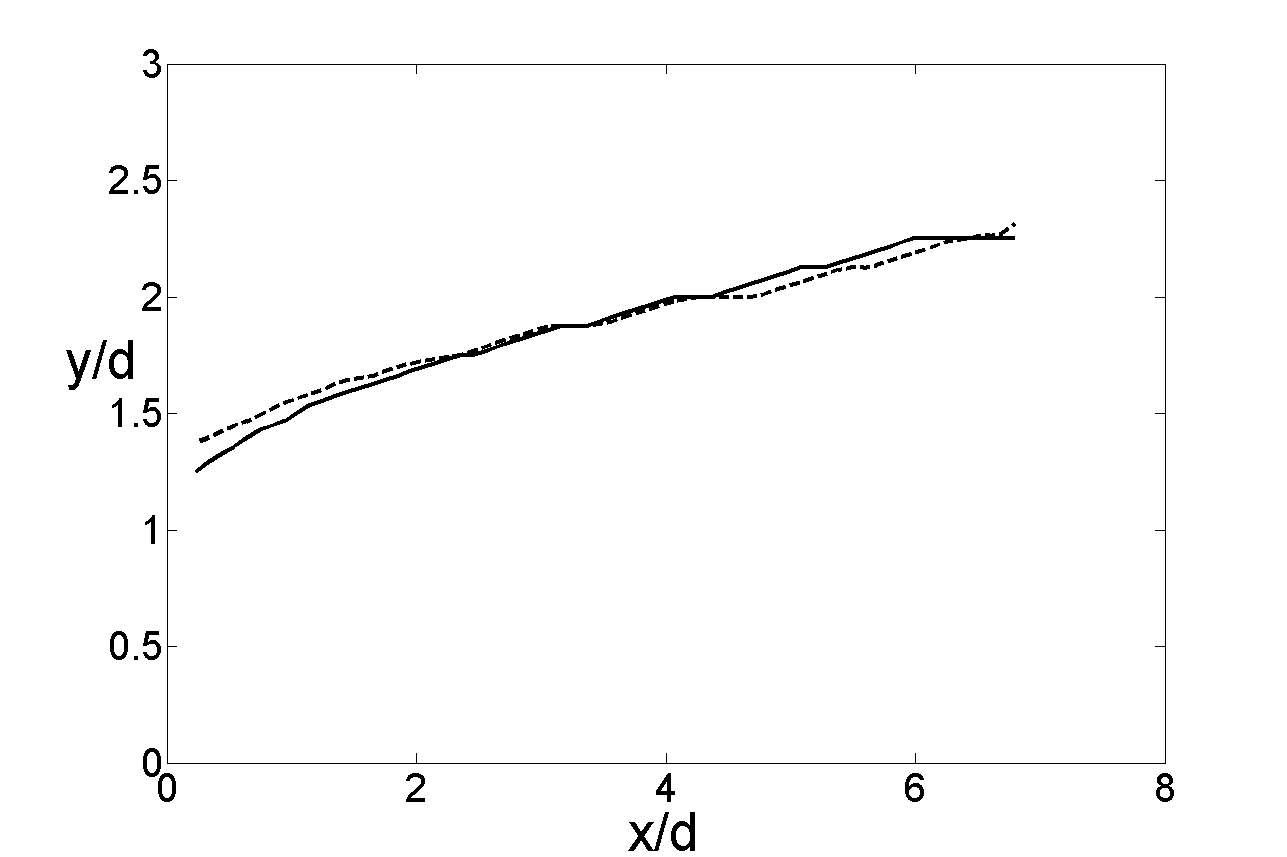} & \includegraphics[width=0.5\textwidth]{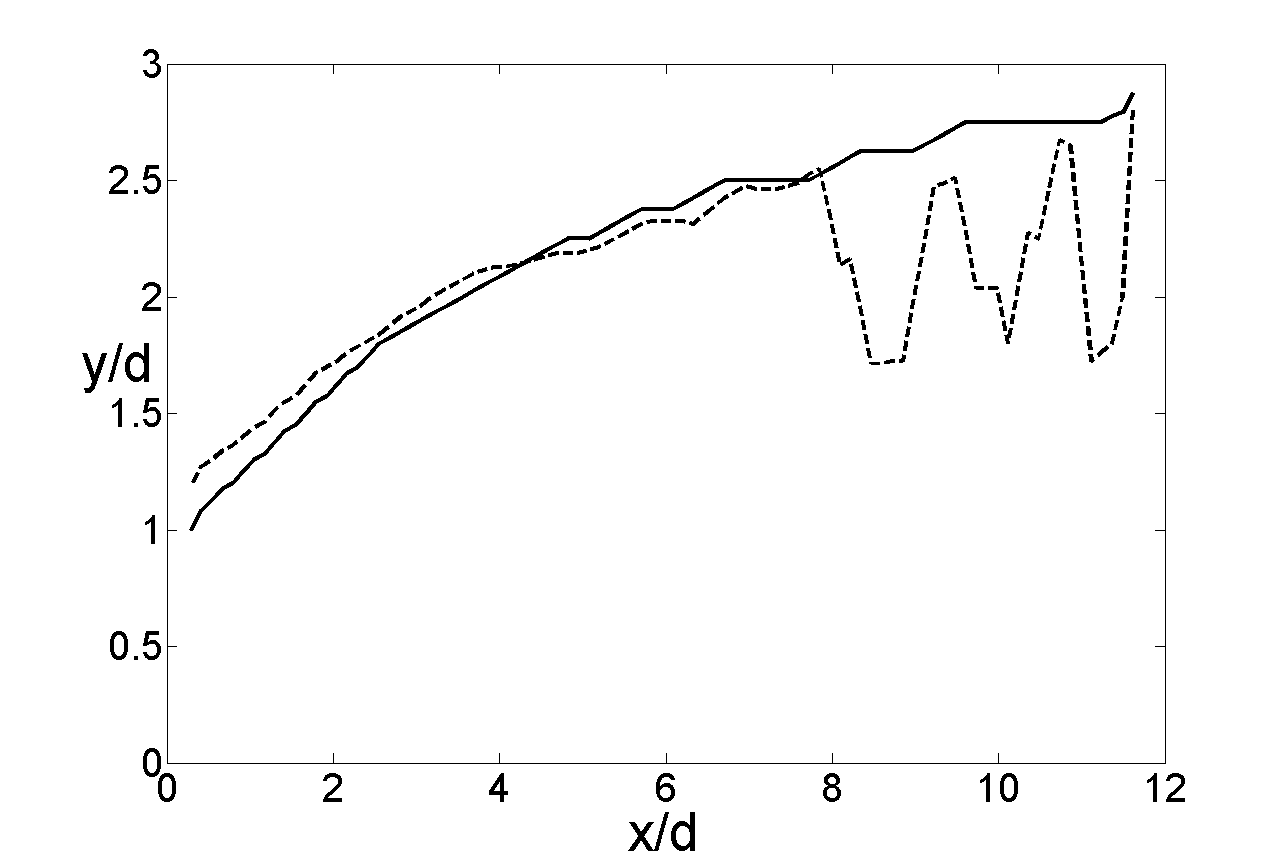}\\
a) & b)
\end{tabular}
\end{center}
\caption{a) CVP trajectories computed using vertical velocity (-) and $\lambda_{ci}$ (- -) for configuration 3. b) CVP trajectories computed using vertical velocity (-) and $\lambda_{ci}$ (- -) for configuration 2.}
\label{fig:VsLci}
\end{figure}

  The method for CVP trajectory computation featured above is not self-sufficient as it does not yield the start of the trajectory. 
To determine where to start the trajectory we track the vortex pair, by computing the two maxima of swirling strength $\lambda_{ci}$ in every constant cross section. This allows us to compute the intensity of the vortex pair $I_{CVP}$ along the trajectory of its cores. Fig. \ref{fig:JetvsCVPtraj} b shows the intensity of the strongest core for configuration 10. We define the start of the CVP trajectory as the abscissa of the maximum of swirling strength of the strongest vortex core which corresponds to lateral shear on the side on the jet.
\\
It might seem unduly complicated to track the outflow instead of the vortex cores themselves. Indeed another way of defining the CVP trajectory is by computing the mean trajectory of both streamwise vortex cores, however it is not as practical. Fig. \ref{fig:VsLci} a and \ref{fig:VsLci} b show CVP trajectories for configuration 10 computed by both methods. Fig. \ref{fig:VsLci} a shows that CVP trajectories computed using vertical velocity and $\lambda_{ci}$ are in good agreement.This demonstrates the relevance of detecting the CVP using the outflow. Fig. \ref{fig:VsLci} b shows the (more common) case where difference in the strength of the vortex cores induces large fluctuations in computed trajectory using swirling strength. Similarly, tracking only one vortex core is much less reliable.
Trajectories extracted with $\lambda_{ci}$  are less reliable, specifically when the intensity of the vortices differs. For our experimental data we obtain considerably better results when considering the locus of vertical velocity maxima than for the locus of $\lambda_{Ci}$ maxima.

\begin{figure}
\begin{center}
\begin{tabular}{c c}
\includegraphics[width=0.5\textwidth]{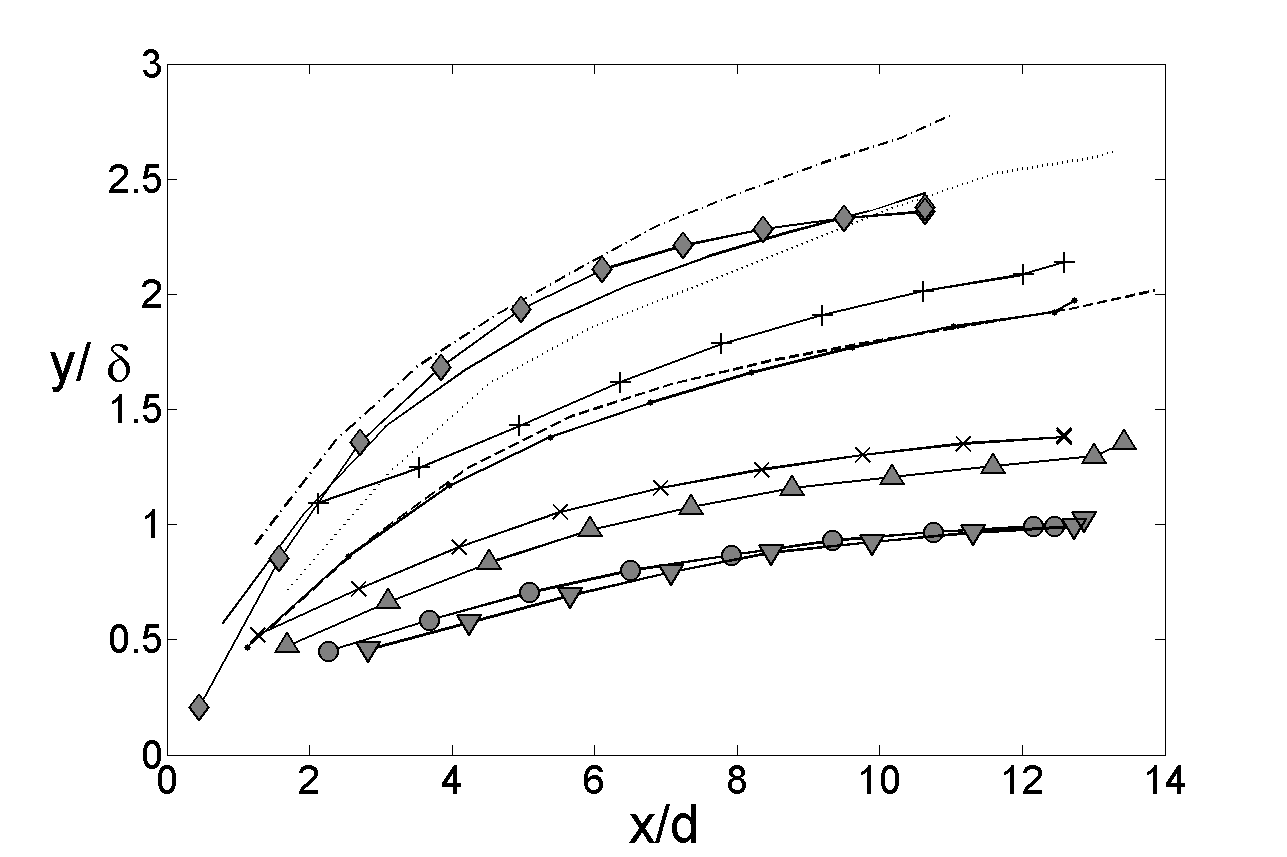}  & \includegraphics[width=0.5\textwidth]{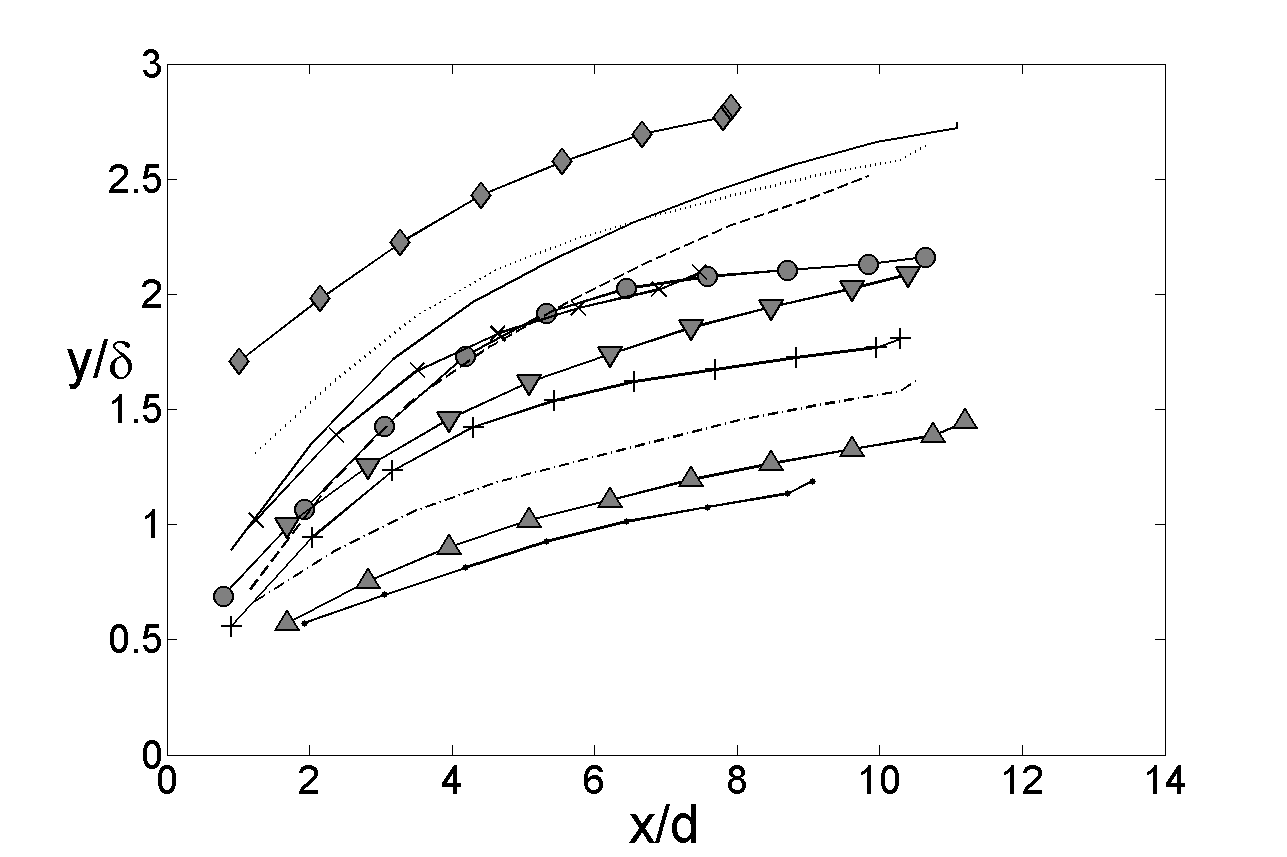} \\
a) & b)
\end{tabular}
\end{center}
\caption{  a ) Trajectories for configurations 1 to 11. b) Trajectories for configurations 12 to 22. Markers are detailed in table \ref{tab:configs}.}
\label{fig:alltraj}
\end{figure}

Trajectories were computed in a volume, but they are very close to the symmetry plane. Therefore only the y-component of the trajectory will be analyzed hereafter. We show on Fig. \ref{fig:alltraj} a and b all 22 computed trajectories using non-dimensional coordinates ($y/\delta$, $x/d$). Trajectories are widely distributed inside and outside the boundary layer (between 0.5 to 3.5 $\delta$).

\section{Definition and relevance of the momentum ratio $r_m$}
\label{sec:9}
In most of the previous studies of JICF, the velocity ratio $r$ is considered as the key parameter despite its limitations: it does not take into account some important features such as the boundary layers of the jet and the cross-flow. Indeed, \cite{Muppidi2005} have shown that the classic $rd$ scaling was not sufficient to collapse all jet trajectories published in the literature onto a single curve. They suggest that the jet exit velocity profile as well as the cross-flow boundary layer thickness influence the jet. This is supported by the analysis of the influence of jet exit velocity profile on jet trajectories conducted by \cite{New2006}. To account for momentum distribution in the jet and boundary layer we introduce a momentum ratio $r_m$ integrating the momentum distribution of the jet and cross-flow boundary layer (also mentioned in \cite{Muppidi2005}), equation \ref{eq:rm1}:

\begin{equation}
 r_m^2=\frac{\frac{1}{S} \int_S V_j^2 dS} { \frac{1}{\delta} \int_0^{\delta} U_{cf}^2 dy}
\label{eq:rm1}
\end{equation}

where $U_{cf}(y)$ is the cross-flow velocity at $y$ and $S$ is the jet nozzle exit section. To highlight the difference with the velocity ratio $r$, $r_m$ can be decomposed in three parts:

\begin{equation}
r_m=(\sqrt{r_{m,jet}}\cdot \frac{1}{\sqrt{r_{m,cf}}}) \cdot r
\label{eq:rm2}
\end{equation}

with 

\begin{eqnarray}
r_{m,jet}=\frac{S \int_S V_j^2 dS}{(\int_S V_j dS)^2}=  \frac{\overline{V_j^2}}{\overline{V_j}^2},  &&
r_{m,cf}=\frac{\int_0^{\delta} U_{cf}^2 dy}{\delta U_{\infty}^2}=  \int_0^{1} (\frac{U_{cf}}{U_{\infty}})^2(\frac{a}{\delta}) da
\end{eqnarray}

\begin{figure}
  \begin{center}
\begin{tabular}{c c}
\includegraphics[width=0.5\textwidth]{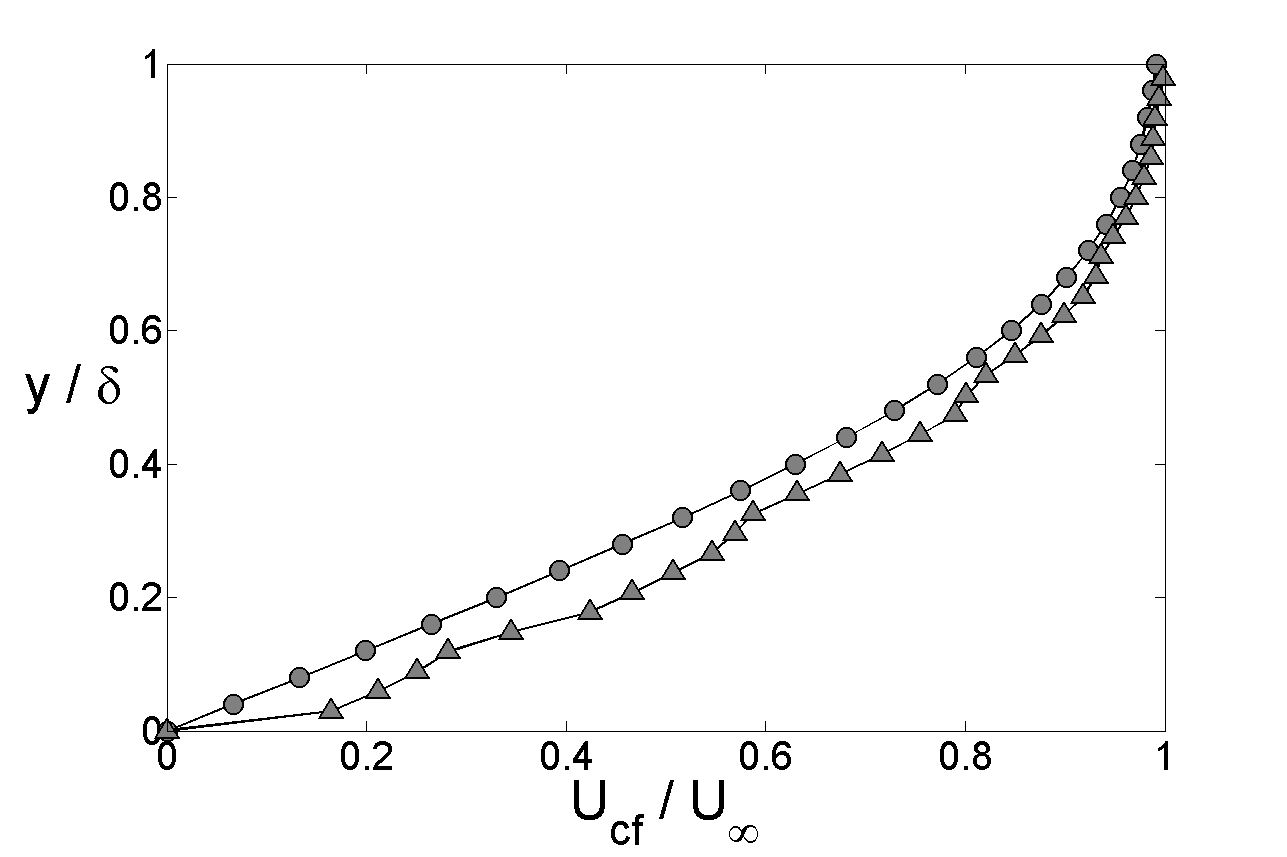} & \includegraphics[width=0.5\textwidth]{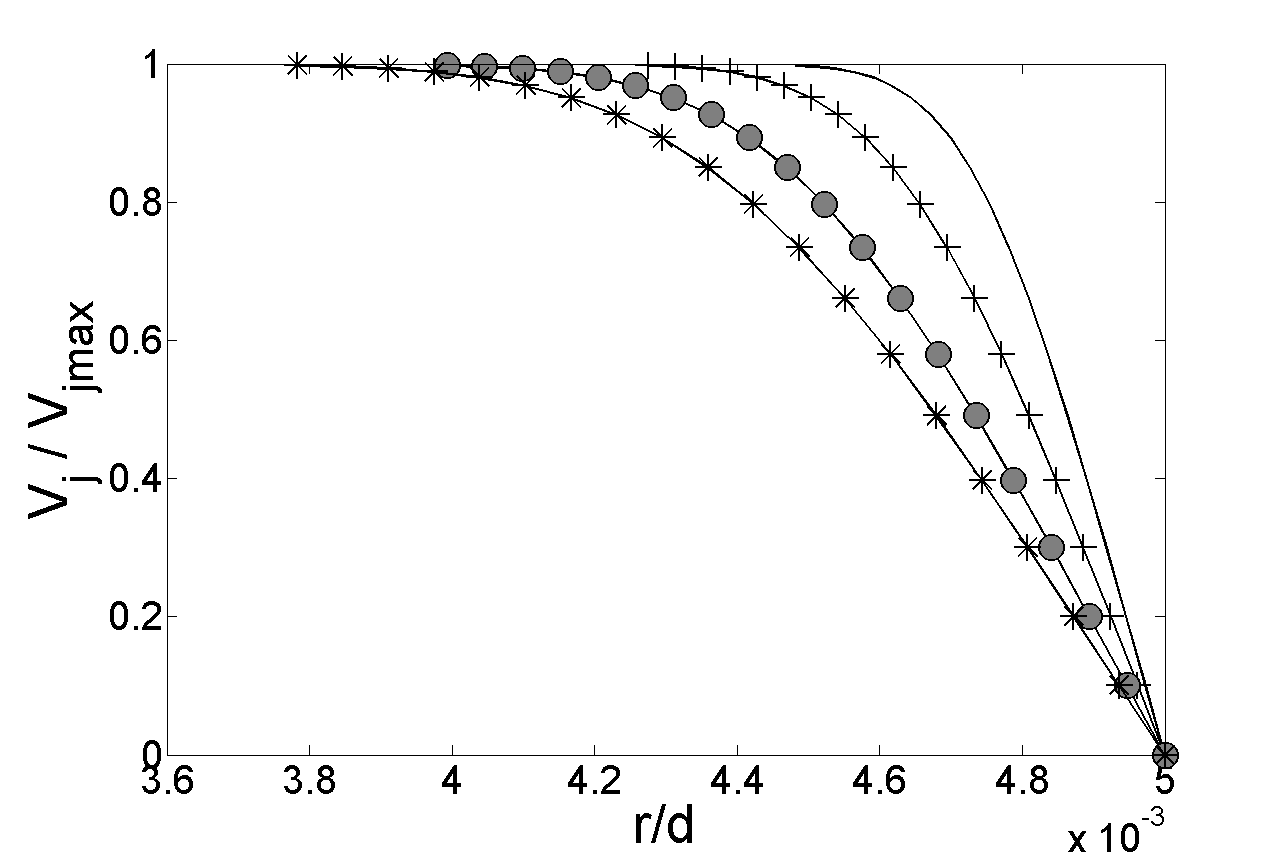} \\
a) & b)

\end{tabular}
\caption{(a)  \FilledTriangleUp : Experimental boundary layer velocity profile for configuration 13, \FilledCircle : theoretical Blasius boundary layer profile.
b) Theoretical jet velocity profiles as a function of injection length. $l_n=0.5cm, r_{m,jet}=1.04 $ ( $-$ ); $l_n=1cm, r_{m,jet}=1.05$ ( $+$ );  $l_n=2cm, r_{m,jet}=1.07$ ( \FilledCircle ); $l_n=3cm, r_{m,jet}=1.09$ ( $-+$ ) }
\label{fig:rm}
  \end{center}
\end{figure}

This decomposition involves two non-dimensional shape factors: $r_{m,cf}$ and $r_{m,jet}$. $r_{m,cf}$ accounts for the momentum distribution in the cross-flow boundary layer ($0<r_{m,cf}<1$, by definition), while $r_{m,jet}$ accounts for the momentum distribution in the jet.\\
 To quantify the influence of the velocity profiles on these two new shape factors, we use the boundary layer velocity profiles shown on Fig. \ref{fig:rm} a for the cross-flow and the velocity profiles shown on Fig. \ref{fig:rm} b for the jet. Typical values obtained for $r_{m,cf}$ with the Blasius ($r_{m,cf} = 0.52$) or experimental ($r_{m,cf} = 0.57$) boundary layer profiles are shown on  table \ref{tab:profiles}. $r_{m,cf} \approx 1$ corresponds to a plug profile. This is coherent with the fact a boundary layer with much momentum near the wall  leads to a lower trajectory. In the following, the value of $r_{m,cf}$  is computed using experimental velocity data.

\begin{table}[H]
  \begin{center}

  \begin{tabular}{|c | c c |c| c | c c |}
  \hhline{|-|--|~|-|--|}
profiles  & Blasius& Experimental & &profiles & Plug / Tophat  & Parabolic\\ \hhline{|-|--|~|-|--|}
$r_{m,cf}$  & $0.52$& $0.57$  && $r_{m,jet} $ & $1$ & $1.33 $\\ \hhline{|-|--|~|-|--|}

  \end{tabular}
  \caption{Significant values for $r_{m,cf}$ and $r_{m,jet}$ for typical velocity profiles.}
  \label{tab:profiles}
  \end{center}
\end{table}

Our measurement method does not allow for a sufficient resolution of the velocity profiles at the exit of the jet nozzle to satisfactorily compute the value of $r_{m,jet}$ with experimental data.  Consequently $r_{m,jet}$ is estimated using the expression for boundary layer thickness in a smooth pipe proposed by \cite{Mohanty1978}. Knowing the jet velocity and the nozzle injection length we compute the analytical jet exit velocity profiles shown in Fig. \ref{fig:rm} b, before computing the associated values for $r_{m,jet}$. Values for $r_{m,jet}$ vary between $1$ (for a top-hat profile) and $1.33$ (for a parabolic profile). \cite{Muppidi2005} show that a parabolic JICF achieves higher penetration than a top-hat JICF. This is corroborated by experimental work by \cite{New2006}. Their interpretation is that the thicker shear layers associated with parabolic JICF delay the formation of leading-edge and lee-side vortices. Therefore possible values taken by $r_{m,jet}$ are coherent with the effect of jet velocity profile on jet trajectory.  Consider two jet trajectories with identical velocity ratios, boundary layer profiles and jet exit diameter but with different exit velocity profiles: one with a parabolic profile, one with a top-hat velocity profile. The parabolic jet penetrates deeper resulting in a higher overall trajectory. For both cases, values of $r$ are identical. Values of $r_m$  are different, making $r_m$ the more relevant parameter.
\\
As shown in table \ref{tab:configs}, we obtain $0.75<r_{m}<4.10$ corresponding to $0.55<r_{m,cf}<0.67$ and $1.05<r_{m,jet}<1.13$ for our configurations.

\section{Influence of experimental parameters on CVP trajectories }
\label{sec:10}

\subsection{Influence of velocity ratio and boundary layer thickness}
\label{sec:11}

\begin{figure}[h]
\centering
\begin{tabular}{c  c}
\includegraphics[width=0.5\textwidth]{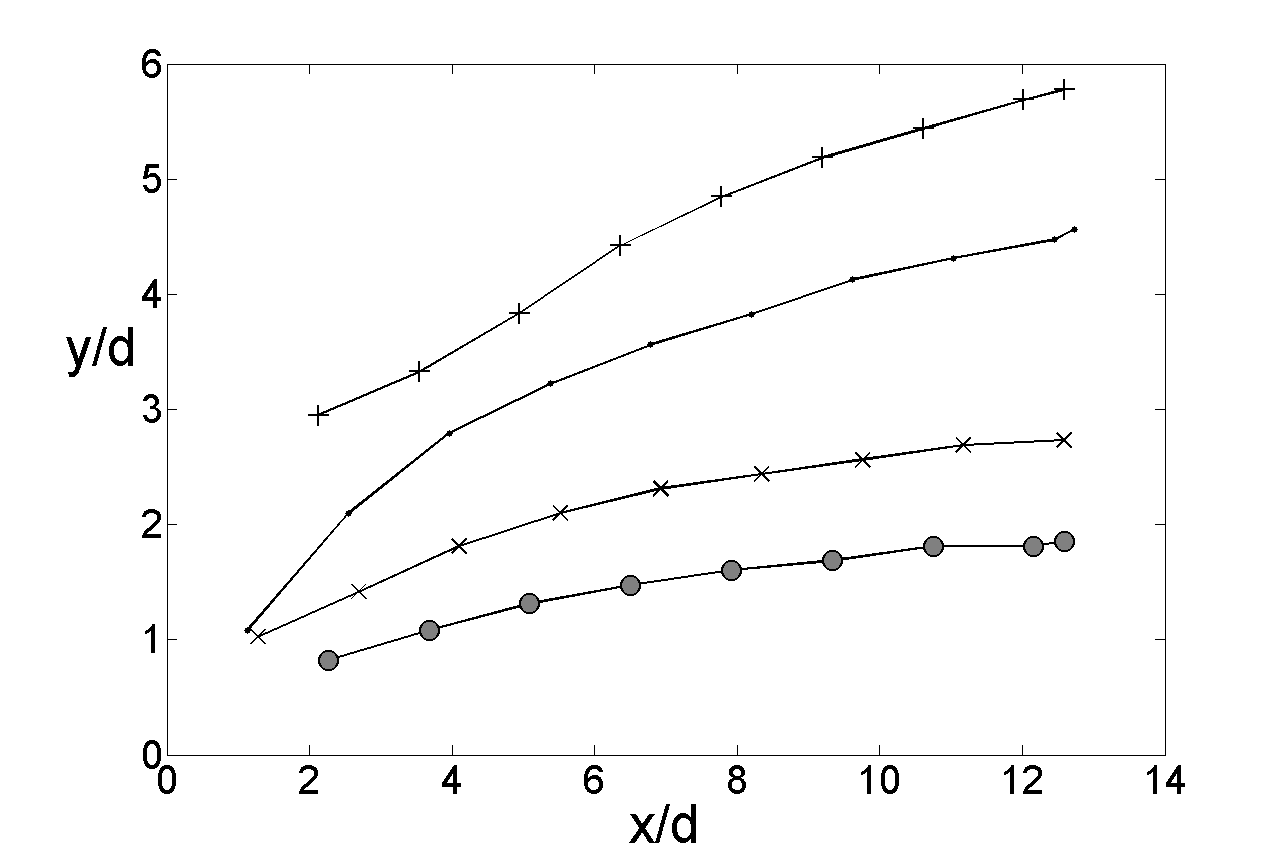} & \includegraphics[width=0.5\textwidth]{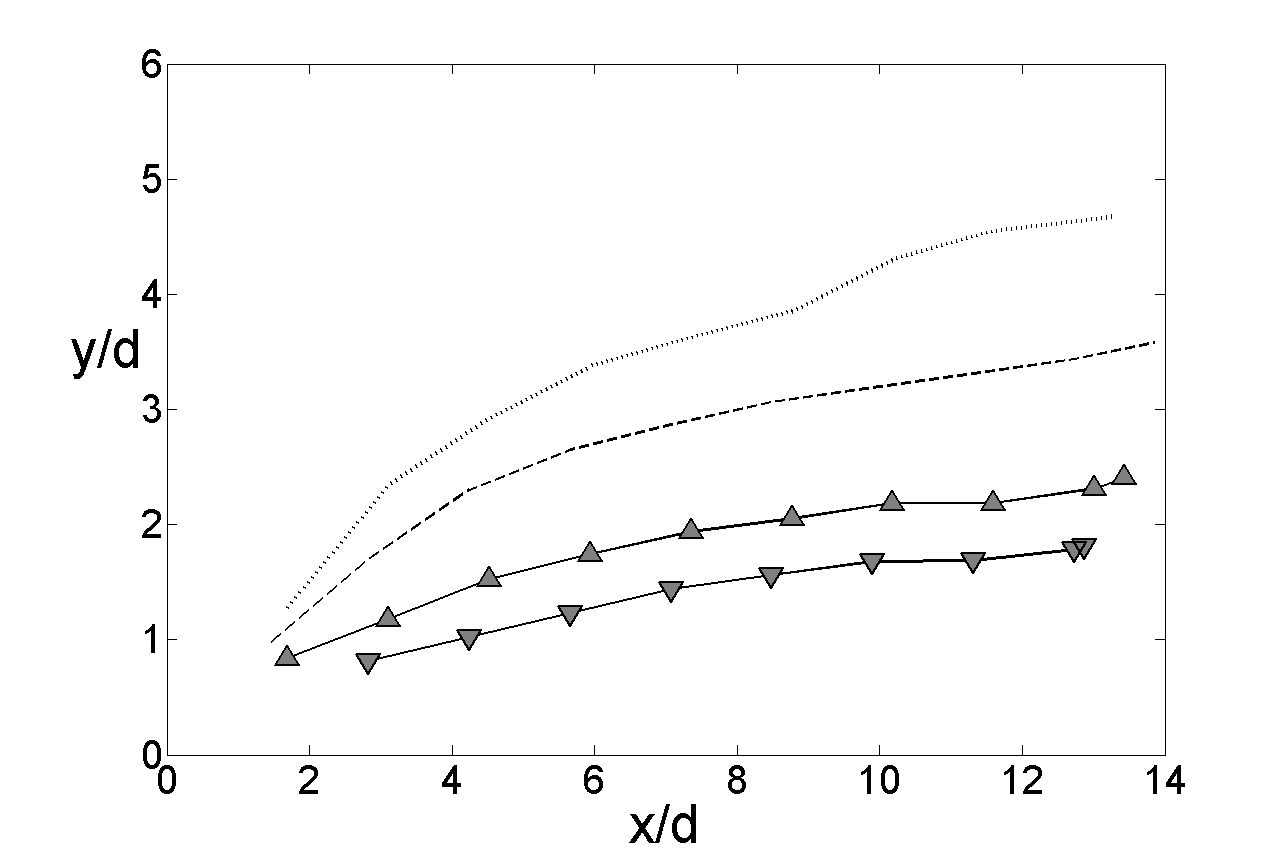} \\
a) & b)
\end{tabular}
\caption{   a) Influence of the velocity ratio on the CVP trajectories with  constant jet exit velocity for $r=0.54$ (-\FilledCircle), $r=0.74$ ($-\times$), $r=1.14$ ($-\bullet$), $r=1.62$ ($-+$). b) Influence of the velocity ratio on the CVP trajectories with constant boundary layer thickness and profile for $r=0.51$ (-\FilledTriangleDown), $r=0.83$ (-\FilledTriangleUp), $r=1.07$ ($- -$), $r=1.24$ ($\cdot \cdot \cdot$).} 
\label{fig:trajsr}

\end{figure}

\begin{figure}[h]
\centering
\includegraphics[width=0.5\textwidth]{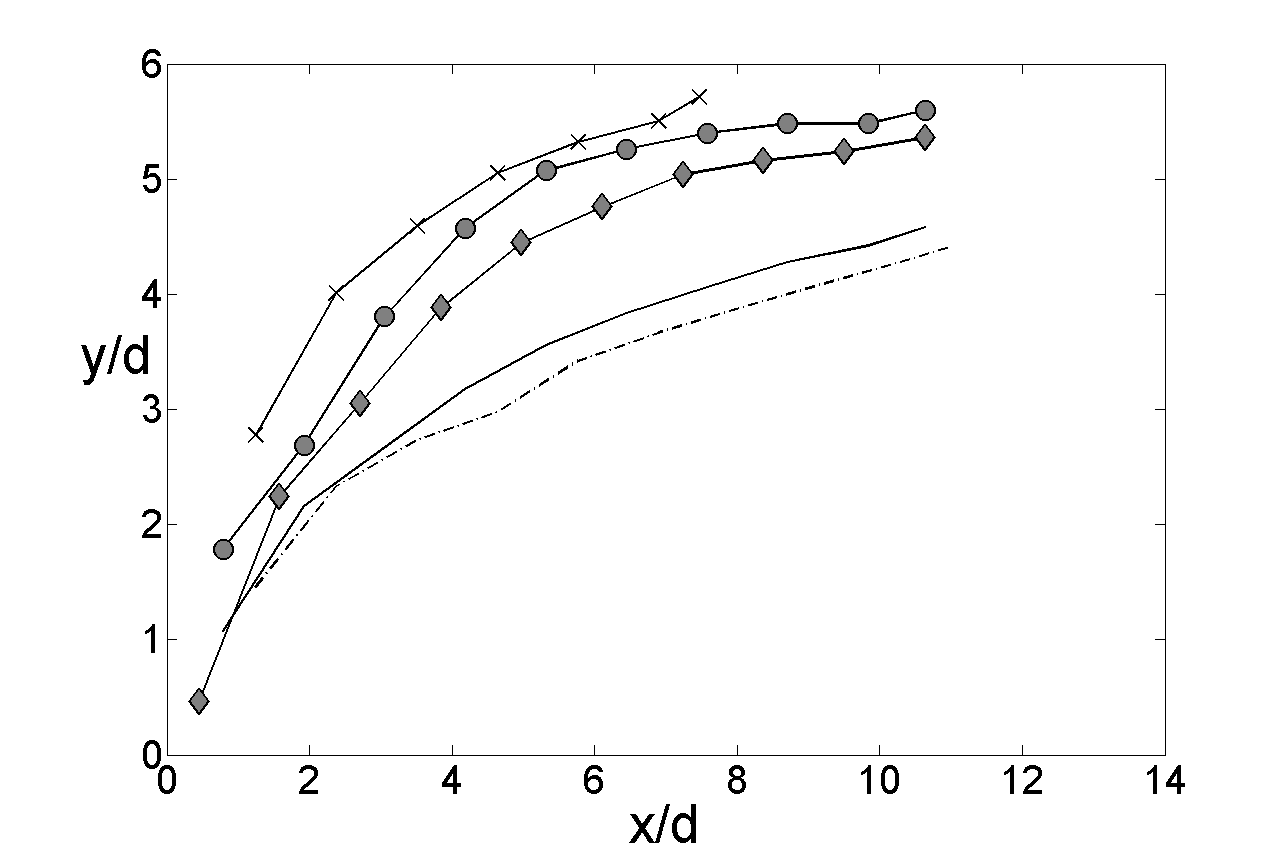}
\caption{ Influence of boundary layer thickness on CVP trajectories:  $\delta=1.59d$ ($- \cdot$), $\delta=1.87d$ ($-$), $\delta=2.25d$ (-\FilledDiamondshape), $\delta=2.59d$ (-\FilledCircle), $\delta=2.79d$ ($-\times$).}
\label{fig:trajsdelta}

\end{figure}

 Fig. \ref{fig:trajsr} a and b show the influence of velocity ratio.  The x and y coordinates are scaled by $d$.
\\
In Fig. \ref{fig:trajsr} a the velocity ratio ranges from $r=0.54$ to  $r=1.62$, while jet exit velocity and profile are kept constant. Cross-flow velocity changes and therefore boundary layer thickness changes also.\\
  In Fig. \ref{fig:trajsr} b velocity ratio ranges from $r=0.51$ to $r=0.84$, while cross-flow velocity, boundary layer thickness and profile are kept constant. Jet exit velocity and profile change. It should be noted that with a constant injection length it is experimentally impossible to vary jet velocity ratio while keeping jet exit velocity profile and boundary layer thickness constant.  In all cases the trajectory of the CVP rises with an increase in velocity ratio.
\\\\
Fig. \ref{fig:trajsdelta} compares CVP trajectories for different values of the boundary layer thickness. All other parameters being equal CVP trajectories penetrate deeper when the cross-flow boundary layer is thicker. The same result has been obtained numerically for jet trajectories by \cite{Muppidi2005}   and observed by \cite{Cortelezzi2001}. This is explained by the fact a thinner boundary layer has more momentum close to the jet exit. Jet trajectories bend earlier and the resulting CVP is created closer to the wall, thus resulting in an overall lower CVP trajectory.

\subsection{Influence of jet exit velocity profile through variation of injection length}

\begin{figure}[h]
\begin{center}
\begin{tabular}{c  c}
\includegraphics[width=0.5\textwidth]{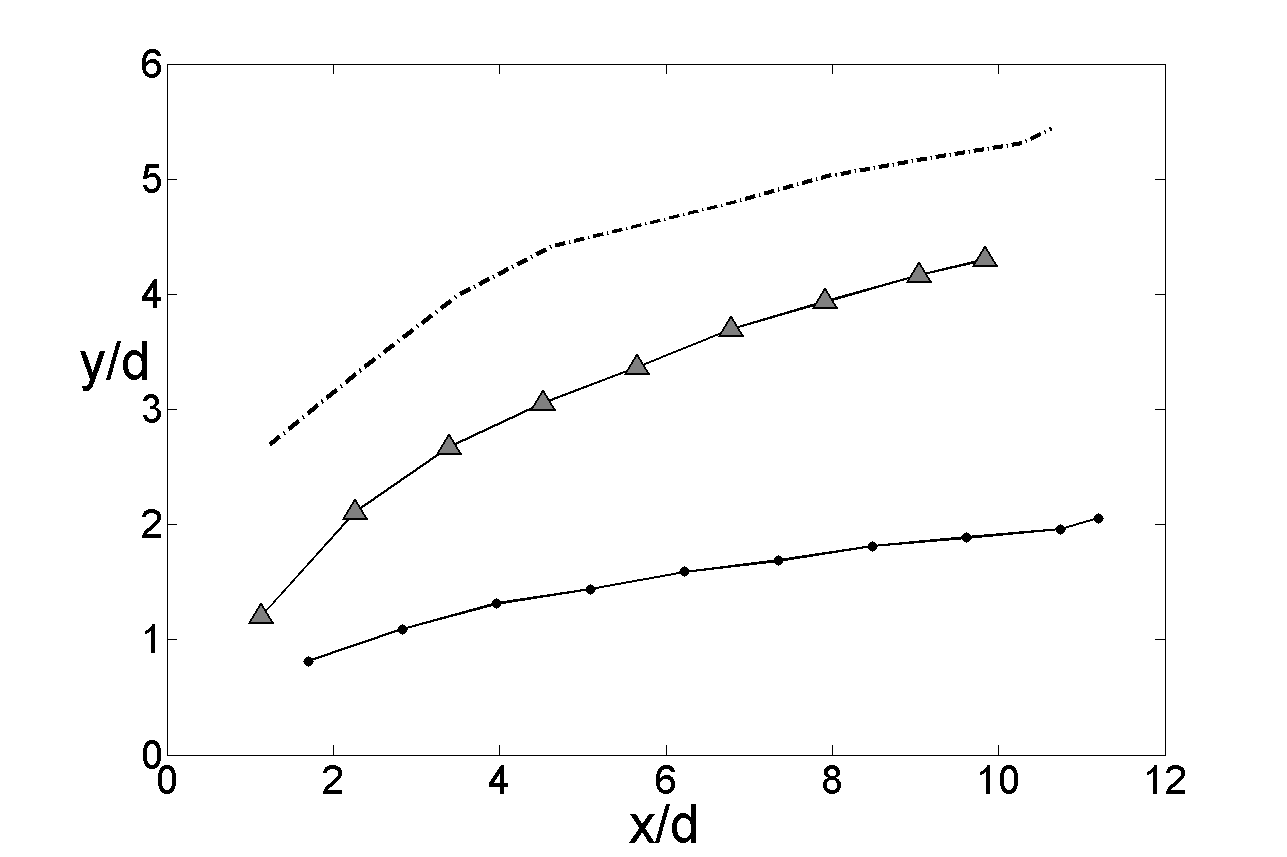} & \includegraphics[width=0.5\textwidth]{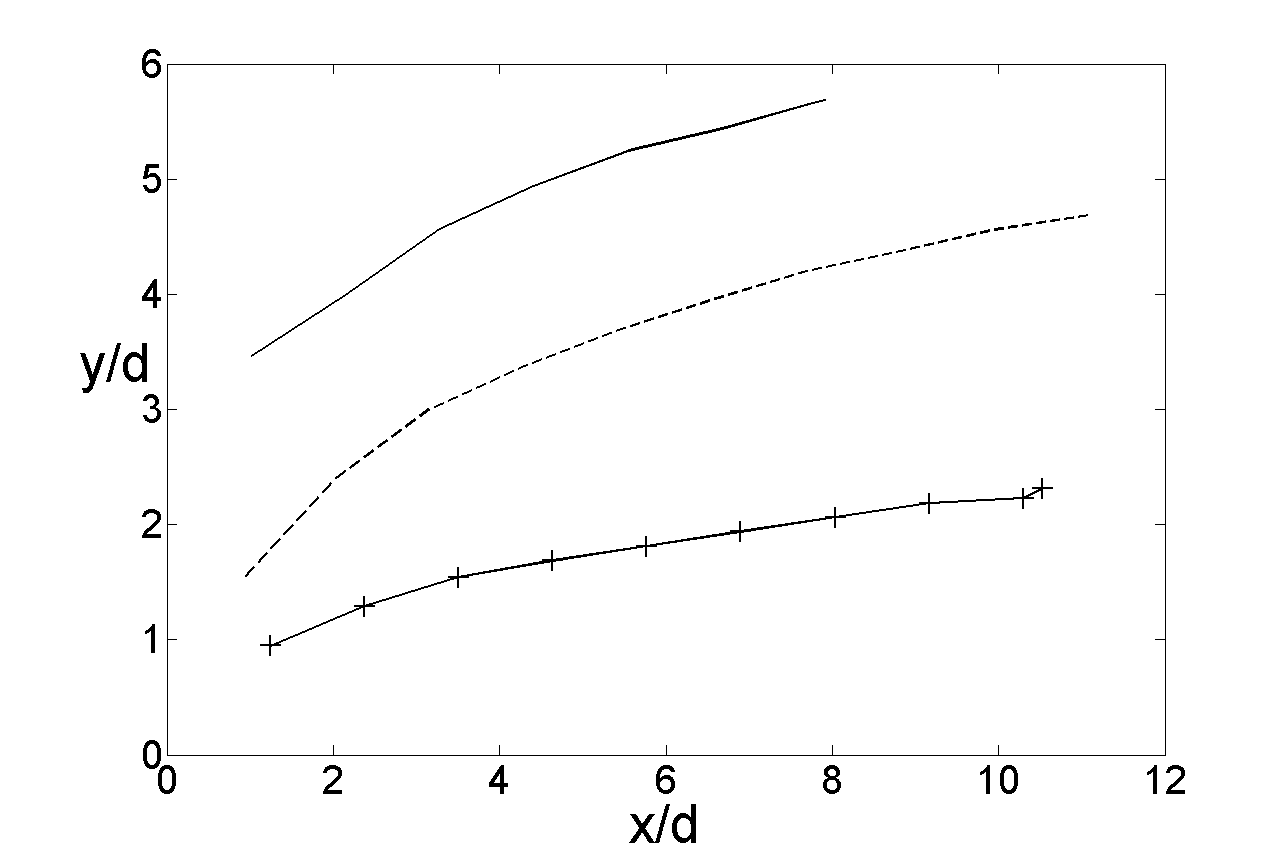}\\ 
a) & b)
\end{tabular}
\caption{a)   Influence of injection length $l_n$ on CVP trajectories for $r=0.96$ and constant boundary layer thickness and profile: $l_n=0.5cm$ ($-\bullet$), $l_n=2cm$ (-\FilledTriangleDown), $l_n=3cm$ ($\cdot \cdot \cdot$). b)  Influence of injection length $l_n$ on CVP trajectories for $r=1.9 $ and constant boundary layer thickness and profile: $l_n=0.5cm$ ($-+$), $l_n=2cm$ ($- -$), $l_n=3cm$ ($- \cdot -$).}
\label{fig:trajInjectionLength}
\end{center}
\end{figure}

For a constant jet flowrate, changing the injection length modifies the jet exit velocity profile. Fig. \ref{fig:trajInjectionLength} shows CVP trajectories for different nozzle injection lengths, while cross-flow velocity and mean jet velocity are kept constant for two different velocity ratios. An increase in injection length leads to more parabolic jet exit velocity profiles as illustrated in Fig. \ref{fig:rm} b. Fig. \ref{fig:trajInjectionLength} shows that the more parabolic the velocity profile, the higher the CVP trajectory. Although nozzle lengths do not come close to what one would need to ensure a full parabolic profile ($l_n>60d$) the effect on CVP trajectory is significant. This is an important result: even a small modification of the exit velocity profile can change the height of the CVP trajectories significantly. This sensitivity could be due to the low velocity ratios featured for this data. This issue is investigated in section \ref{sec:16}. Apart from the discussion on trajectory scaling, this data clearly illustrates how it is possible to obtain higher trajectories without spending more energy, only by modifying the design of the injection.

\label{sec:12}

\section{Trajectory Scaling}
\label{sec:13}

\subsection{Scaling quality factor}
\label{sec:14}

In order to quantitatively compare how well different scalings collapse trajectories  we define a non-dimensional scaling quality factor $Q$.
A perfect scaling would collapse all trajectories onto a single curve, in other words the scattering would be null. This can be characterized by a quantitative criterium. \\
For a given abscissa $\tilde{x}$ we define $Y(\tilde{x})$ the set of values taken by the trajectories at this abscissa. We define $\left[\tilde{x}_{start}, \tilde{x}_{end}\right]$ the range where trajectories exist. $\tilde{x}_{start}$ is then the first abscissa where an outflow region can be identified, i.e where the first trajectory starts, while $\tilde{x}_{end}$ corresponds to where the longest trajectory ends. This range may change depending on how the abscissa is scaled.\\
For our trajectories, the case arises where not all of them are defined for a given $\tilde{x}$. In order to take this into account we introduce $N(\tilde{x})$ and $N_{curves}$, respectively the number of curves defined at abscissa $\tilde{x}$ and the total number of curves considered for scaling. 
The scaling quality factor is defined as the integral of trajectory scatter relative to the mean over the range where these trajectories exist. 

\begin{equation}
Q=\int_{\tilde{x}_{start}}^{\tilde{x}_{end}} \frac{\sigma(Y)}{\overline{Y}}(\tilde{x}).   \frac{N(\tilde{x})}{N_{curves}} d \tilde{x}
\label{eq:Q}
\end{equation}
where $\sigma{(Y)}$ is the standard deviation of Y and $\overline{Y}$ is the mean of Y for a given abscissa $\tilde{x}$. $Q=0$ corresponds to a perfect scaling.
\\
 To take into account the fact that trajectories are not defined over the same spatial range, this relative scatter is weighted by the ratio, $\frac{N(\tilde{x})}{N_{curves}}$. This is done to give more meaning to the collapse of many trajectories than to the collapse of a few. For a set of trajectories defined over the same domain the weight is one, and the definition for $Q$ can be simplified to the expression shown in equation \ref{eq:Qsimple}:
\begin{equation}
  Q=\int_{\tilde{x}_{start}}^{\tilde{x}_{end}} \frac{\sigma(Y)}{\overline{Y}}(\tilde{x}). d \tilde{x}
\label{eq:Qsimple}
\end{equation}
 Normalizing by the mean  is necessary to ensure that multiplication of all trajectories by any constant does not change the value of $Q$. 
This method is applicable to any collection of 2D curves, for any scaling of the x-coordinate. Particularly $Q$ can be used to gauge the efficacy of a given scaling of CVP or jet trajectories. \\
 For clarity, $Q$ is normalized by its value $Q_0$ taken when the data is not scaled, both in $x$ and in $y$.

\subsection{Reflexions on previously published jet trajectories}
\label{sec:15}
\begin{figure}[h]
\begin{tabular}{c c}
 \includegraphics[width=0.5\textwidth]{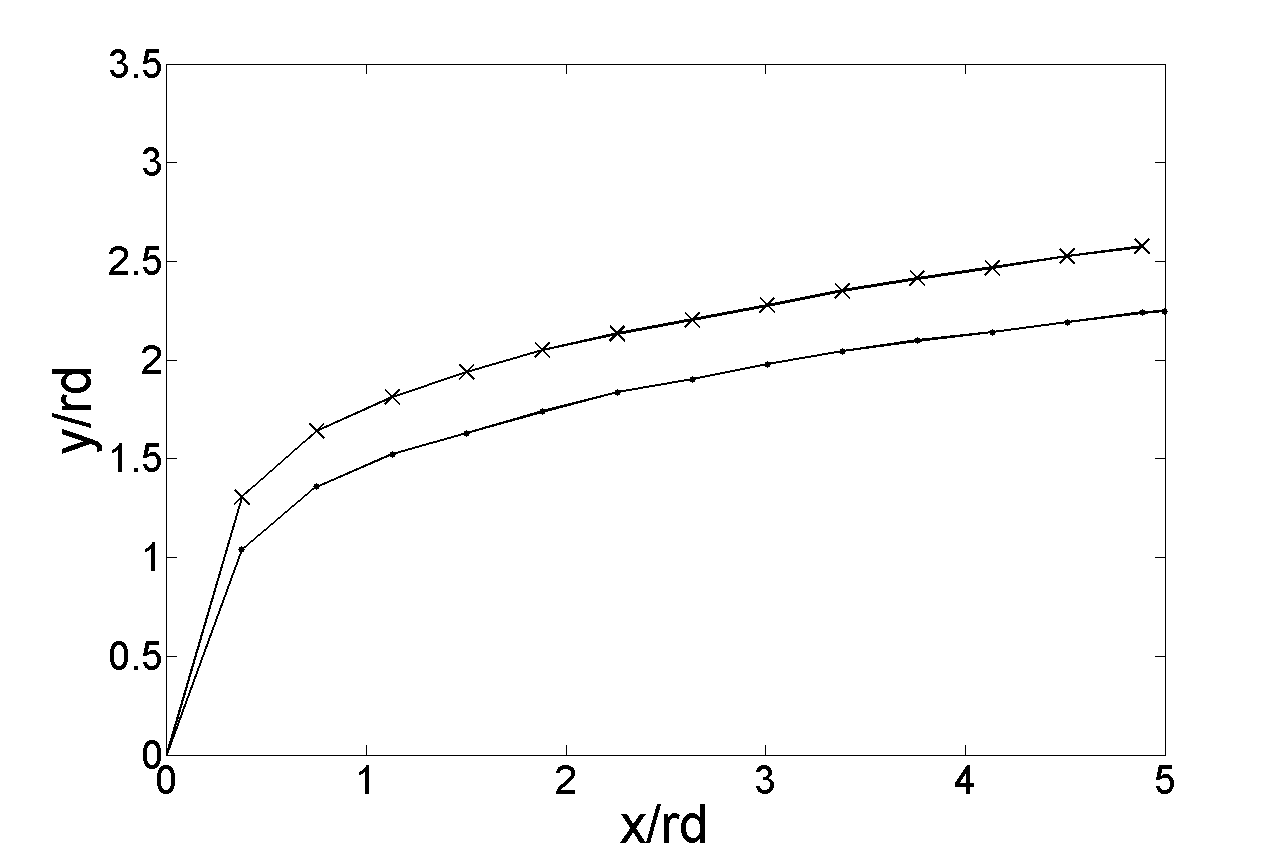} &  \includegraphics[width=0.5\textwidth]{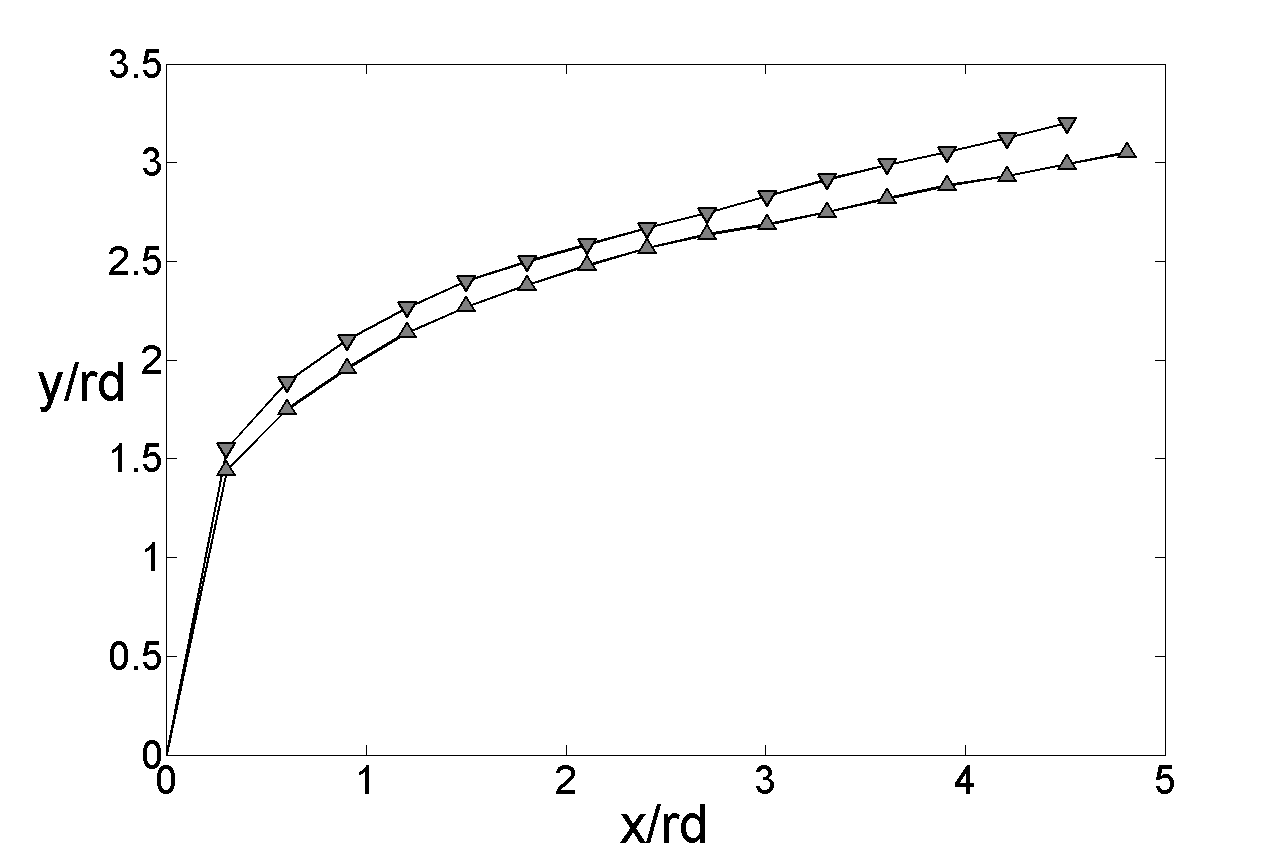} \\
a) & b)
\end{tabular}
\caption{a) Jet trajectories for constant jet exit velocity profile and velocity ratio but varying boundary layer thickness$-\bullet$, $\delta=0.44d$, $-\times$, $\delta=1.32d$. b) Trajectories for constant velocity ratio, boundary layer thickness. Jet exit velocity profile varies between tophat and parabolic. Parabolic : -\FilledTriangleUp,$r_m=8.6$. Top-hat : -\FilledTriangleDown,$r_m=9.16$ . From \cite{Muppidi2005} data.}
\label{fig:MuppidiR}
\end{figure}

To the best of the authors knowledge there are no CVP trajectory data for which jet exit velocity profile, boundary layer thickness and profile are available. However since CVP trajectories  follow the same trends as jet trajectories  (e.g. deeper penetration with increase in momentum ratio) we will begin our discussion using jet trajectory data published in \cite{Muppidi2005}. These results were chosen because the varying parameters were boundary layer thickness and jet exit velocity profile for two different velocity ratios. Table  \ref{tab:Muppidiparam} summarizes  the different parameters used by \cite{Muppidi2005} for their study. The corresponding values of $r_{m,jet},r_{m,cf}$ and $r_m$ were computed using the data presented in their paper. \\
The objective is to derive an approach to the scaling of these jet trajectories which can be applied to CVP trajectories. \cite{Muppidi2005} present a scaling that successfully collapses their trajectories. This scaling uses a parameter $h$ extracted from the data as the y-coordinate at a distance $x=0.05d$. Because CVP trajectories do not start at $x=0$, $h$ is not defined and cannot be used for scaling purposes. Moreover our objective was to validate a more general scaling based on experimental parameters.  Thus an alternate scaling was sought.\\
Fig. \ref{fig:MuppidiR} shows the influence of jet velocity profile and boundary layer thickness for different velocity ratios on jet trajectories. Tophat and parabolic jet exit velocity profiles are used. As shown in table \ref{tab:Muppidiparam} values of $r_m$ are higher for the parabolic profile. For the CVP, jet penetration is higher for parabolic velocity profiles and  for thicker boundary layers.

\begin{table}[h]
  \begin{center}
  \begin{tabular}{|c| c c c c c c c c c|}
 \hline
case & I & II & III & IV & V & VI & VII & VIII & IX \\  \hline
Velocity ratio $r$ & 1.52 & 1.52 & 1.52 & 1.52 & 5.7 & 5.7 & 5.7 & 5.7 & 5.7 \\
$\delta_{80 \%}$ & 1.32d & 1.32d & 0.44d & 0.44d &  1.32d & 1.32d & 0.44d & 0.44d & 6.4d\\
 $r_{m,jet} $& 1.33 & 1.185 & 1.33 & 1.185  & 1.185 & 1.33 & 1.185 & 1.33 &1.33 \\
$r_{m,cf} $ &0.52 & 0.52 & 0.52 & 0.52 & 0.52 & 0.52 & 0.52 & 0.52 & 0.52 \\
$r_m$ & 2.44 & 2.29 & 2.44 & 2.29  & 9.16 & 8.60 & 9.16 & 8.60 & 9.16 \\
Markers & -\FilledCircle & $-\times$ & $-*$ & $-\cdot $ & -\FilledTriangleDown & \FilledTriangleUp & $-|$ & $--$ & $-$  \\  \hline
  \end{tabular}
  \caption{Parameters for jet trajectories from \cite{Muppidi2005}, and corresponding values for  $r_{m,jet},r_{m,cf}$ and $r_m$ obtained using their parameters.}
  \label{tab:Muppidiparam}
  \end{center}
\end{table}

\subsection{Scaling of jet trajectory}
\label{sec:16}

Trajectory scaling  of a circular jet in cross-flow has been the subject of much research (\cite{Pratte1967}, \cite{Smith1998}, \cite{Yuan1998}, \cite{Mungal2001}, \cite{Muppidi2005}, \cite{Gutmark2008}), however no scaling is fully satisfactory. Among the most successful scalings, the $rd$ scaling by \cite{Pratte1967} has proven to collapse most experimental trajectories. For $5 < r <35$, they show the collapse of the centerline trajectory with the $rd$ length scale defined as follows:
\begin{equation}
\frac{y}{rd}=A(\frac{x}{rd})^b
\label{eq:rdscaling}
\end{equation}

where $A=2.05$ and $b=0.28$. 
However more recent works by \cite{Muppidi2005} and \cite{New2006} show that this scaling is not satisfactory for flows where boundary layer thickness and jet exit velocity profile vary. Several attempts were made to scale jet trajectories while accounting for these factors (\cite{Muppidi2005}, \cite{Gutmark2008}).\\

A scaling using $r^{\alpha}$ was introduced by \cite{Karagozian1986} for high velocity ratios. Similarly we choose to consider a scaling using $r_m^{\alpha}$ to account for jet exit velocity profile, where $\alpha$ quantifies the influence of momentum ratio $r_m$ and is unknown \textit{a priori}. To account for the influence of the boundary layer thickness we introduce, in a manner analogous to \cite{Muppidi2005} who use  $(\frac{h}{d})^C$ , the non-dimensional parameter $(\frac{\delta}{d})^\beta$, where $\beta$ quantifies the influence of $\delta$ on trajectory. This leads to the new scaling described in equation \ref{eq:Newscl}:

\begin{equation}
\frac{y}{{r_m}^{\alpha} d (\frac{\delta}{d})^\beta}
\label{eq:Newscl}
\end{equation}

\begin{figure}
\begin{tabular}{c c}
 \includegraphics[width=0.5\textwidth]{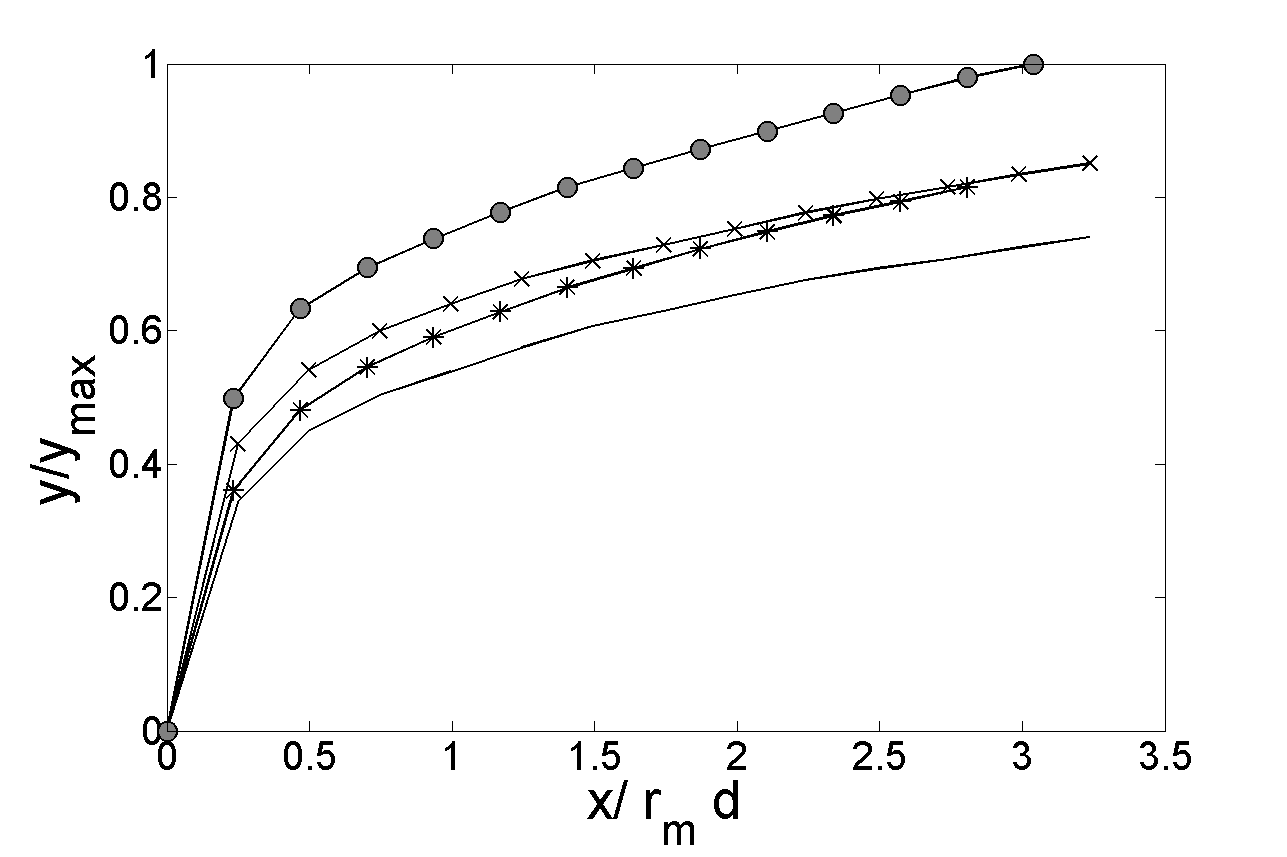} &  \includegraphics[width=0.5\textwidth]{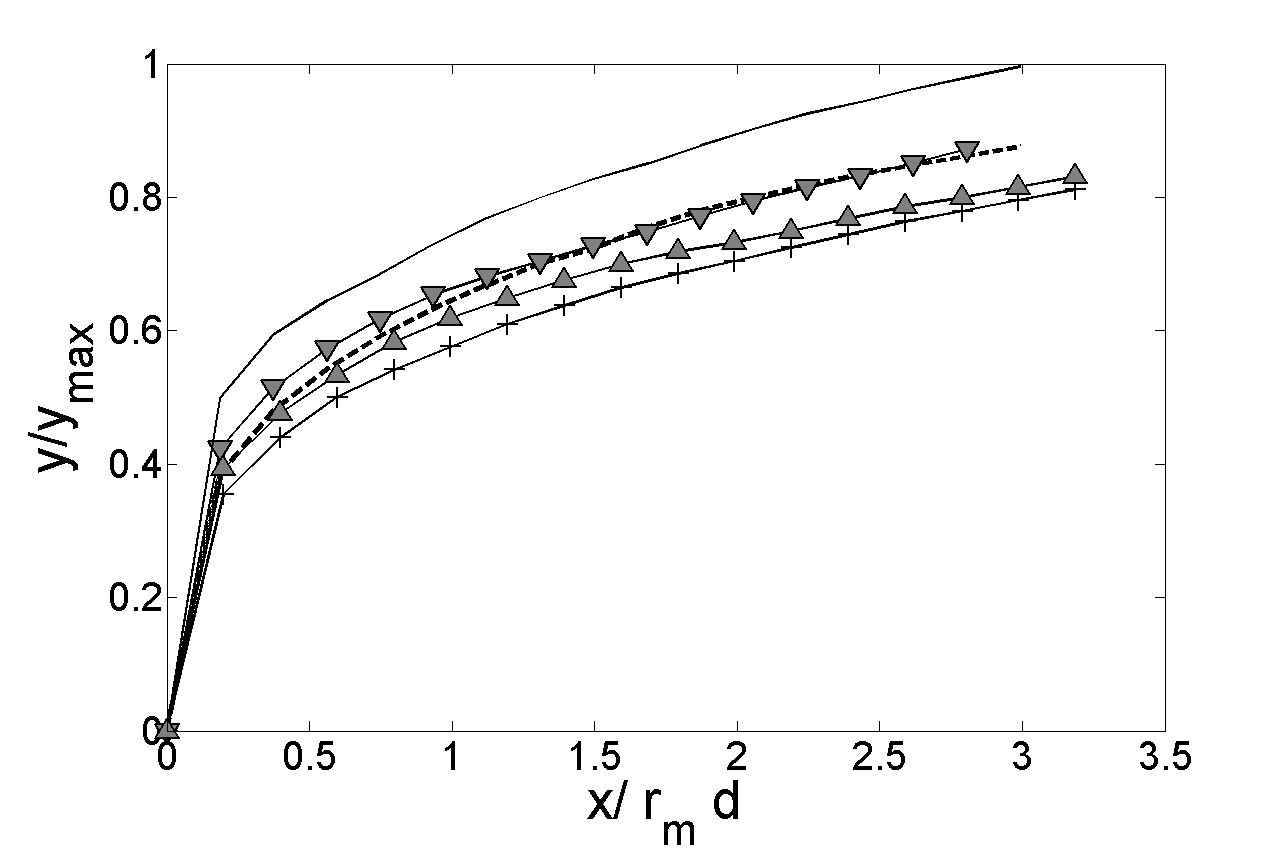} \\
a) & b)
\\
 \includegraphics[width=0.5\textwidth]{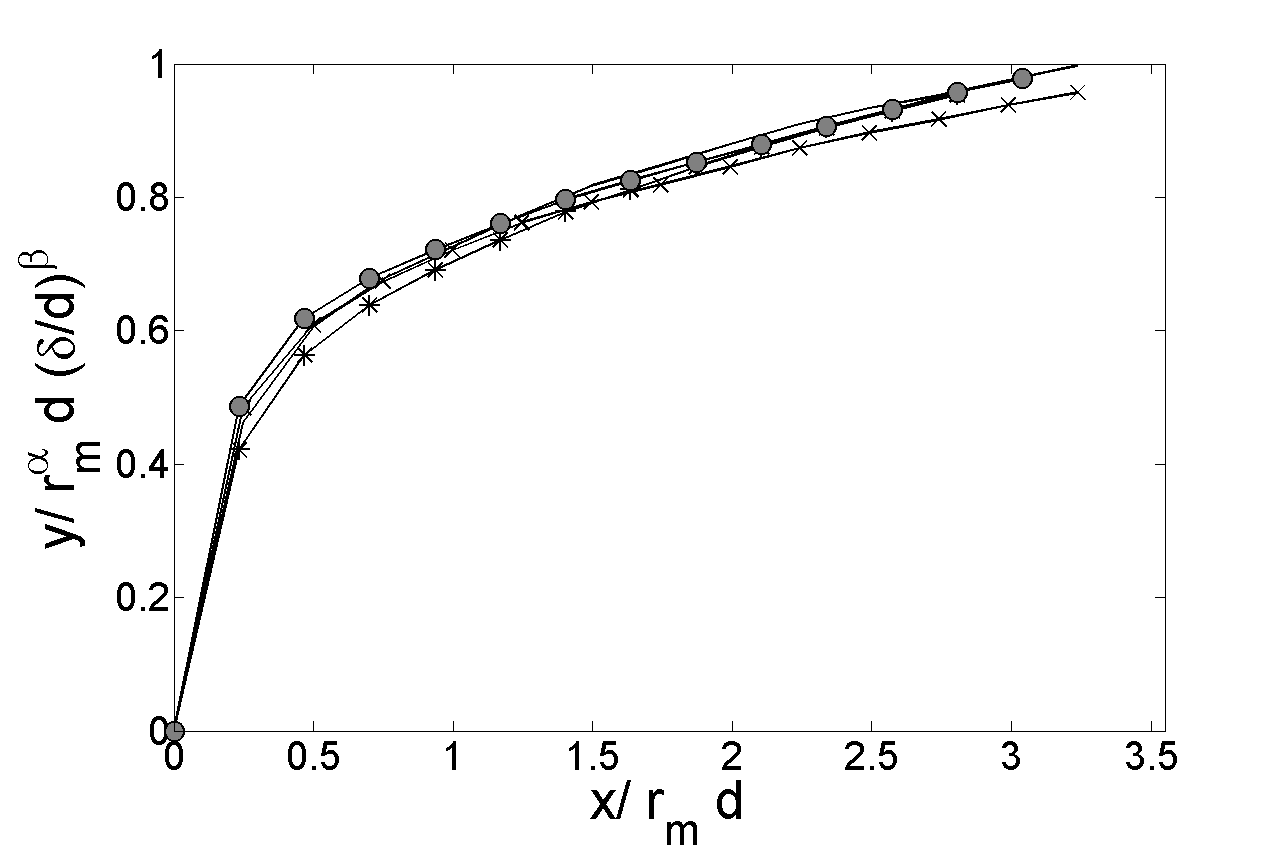}  &  \includegraphics[width=0.5\textwidth]{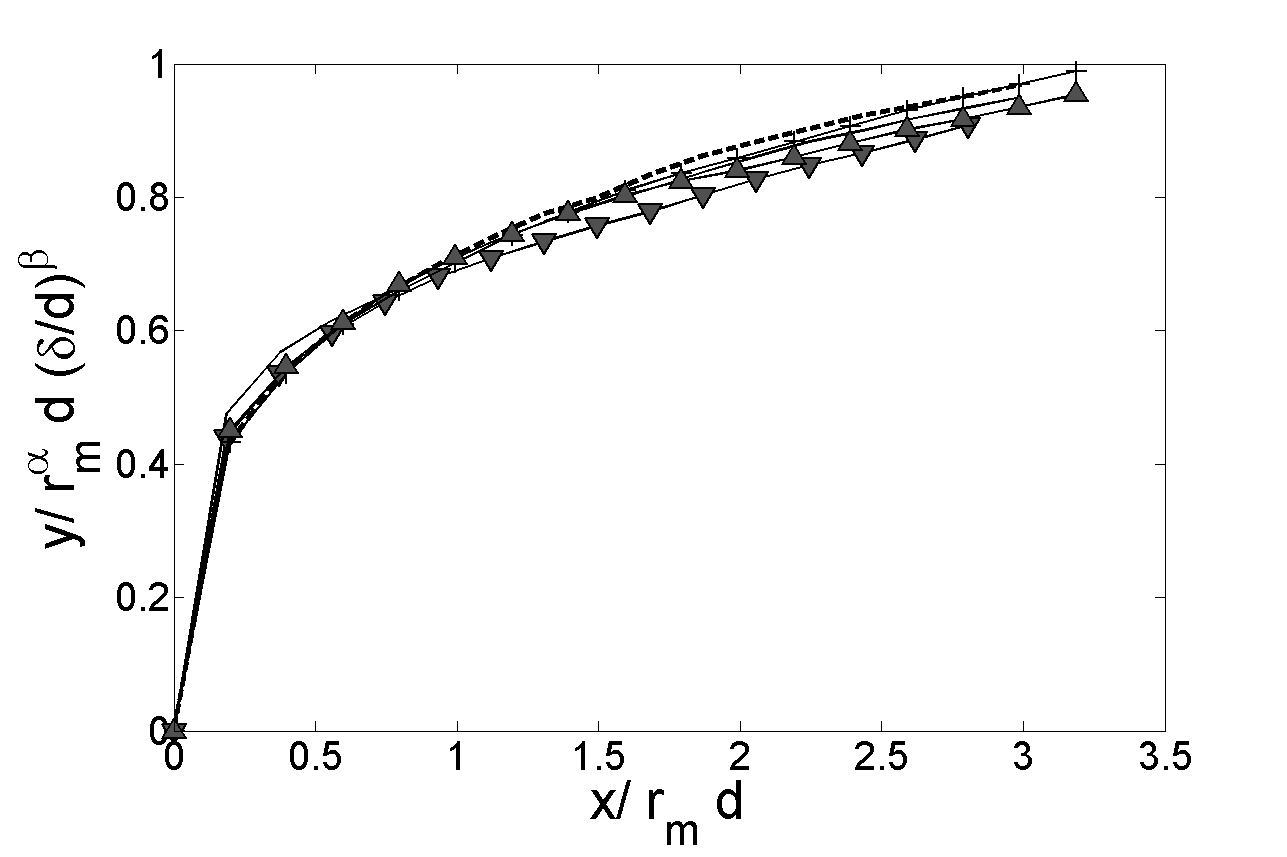} \\
c) & d)
\end{tabular}
\caption{ a) Jet trajectories without scaling in y for  case I (-\FilledCircle), case II ($-\times$),  case III ($-*$),  case IV ($-\bullet$).  b) Jet trajectories  without scaling for case V (-\FilledTriangleDown),  case VI (-\FilledTriangleUp), case VII ( $- |$ ),  case VIII ($--$),  case IX ($-$). c) Jet trajectories from a) with scaling for $r_m^{2.23}$ and $(\frac{\delta}{d})^{0.16}$ leading to $Q=20\%$. d) Jet trajectories from b) with scaling for $r_m^{1.55}$ and  $(\frac{\delta}{d})^{0.06}$ leading to $Q=26\%$. Each set of trajectories is normalized to make the original and scaled trajectory set comparable.}
\label{fig:sclMup}
\end{figure}

Reasoning on the physics of the flow and empirical data, it is possible to define upper and lower bounds for $\beta$ and $\alpha$.\\
 To $\alpha=1,\ \beta=0$ corresponds the scaling $\frac{y}{r_md}$. Having $\beta<0$ would mean the jet penetrates deeper with a decreasing boundary layer thickness, therefore $\beta>0$. Moreover for high velocity ratios where the jet exit profile is usually a plug profile with a fixed boundary layer profile which gives $r_m\propto r$ thus making this scaling equivalent to the $rd$ scaling.\\
 To $\alpha=1,\ \beta=1$ corresponds the scaling $\frac{y}{r_m \delta}$. Having $\beta>1$ would mean deeper jet penetration with decrease in jet diameter, therefore $\beta>1$.  Similarly the data shows how trajectories rise with $d$, therefore $\beta<1$. \\
Using the same reasoning we obtain $\alpha>0$, since trajectories rise with $r_m$. There is however no upper bound on $\alpha$.
\\\\
Fig. \ref{fig:sclMup} {c, d} shows scaled trajectories using equation \ref{eq:Newscl} compared to the classic $rd$ scaling (Fig. \ref{fig:sclMup} {a,b}). For a given set of jet trajectories we search for $\alpha,\beta$ to obtain the best possible collapse. This is equivalent to minimizing the quality factor $Q$, here used in its simplified form defined in equation \ref{eq:Qsimple}.
\\
 Note that jet exit velocity profile and boundary layer thickness do not affect trajectories in the same way for different velocity ratios.\\
 
For $r=5.7$ (Fig. \ref{fig:sclMup} {a,c}), we have $y/(r_m^{1.5} d (\frac{\delta}{d})^{0.05})$ whereas for $r=1.5$ (Fig. \ref{fig:sclMup} {b,d}) we obtain $y/(r_m^{2.3} d (\frac{\delta}{d})^{0.16})$. Indeed two different sets of exponents are found depending on $r$, i.e $\alpha(r)$ and $\beta(r)$. The exponents for $r_m$ and $(\frac{\delta}{d})$ give us insight into how jet trajectory is influenced by jet exit velocity profile and boundary layer thickness. These results indicate that for low velocity ratios, jet trajectory will be more sensitive to variations of the incoming cross flow boundary layer thickness.  While for high velocity ratios boundary layer thickness is less of an issue and the trajectory is mainly influenced by the momentum ratio. The proposed scalings achieve significant collapse as shown in Fig.  \ref{fig:sclMup}.  It also shows how the scaling differs whether high or low velocity ratios are considered.
\\

\begin{table}
  \begin{center}
  \begin{tabular}{|>{\centering}p{2.5cm}|  c |p{4cm} <{\centering}|p{2.8cm} <{\centering}| }
  \hline
Scalings 	                 & No scaling & y/$rd$ &  y/$r_m d$ \\\hline
cases I to IV ($r=1.5$) & $Q=67.8\% $  & $Q=67.8\%$ & $Q=60.8\%$ \\
cases V to IX ($r=5.7$)& $Q=46.0\%$  & $Q=46.0\%$ & $Q=36.6\%$\\
cases I to IX   	      & $Q=100.0\%$  &  $Q=78.4\%$  & $Q=73.3\%$\\
\hline\multicolumn{4}{c}{}\\\hline
Scalings 	                 & Muppidi y/($rd (\frac{h}{d})^C$) & Gutmark $ y/rd /(r_b\cdot(r^2 \frac{d}{\delta})^{0.45})$\newline  $\Leftrightarrow  y/(rd \cdot(r_b(r^2 \frac{d}{\delta})^{0.45}))$ & y/$r_m^{\alpha} d (\frac{\delta}{d})^{\beta}$ \\\hline
cases I to IV ($r=1.5$) & $Q=41.6\%$ &$Q=220.0\%$ & $Q=13.5\%$,\newline  $(\alpha=2.23,\beta=0.16)$\\
cases V to IX ($r=5.7$) & $Q=32.3\%$ &$Q=393.3\%$ & $Q=12.0\%$,\newline  $(\alpha=1.55,\beta=0.05)$\\
cases I to IX   	      & $Q=36.3\% $& $Q=336.9 \%$ & $Q=28.0\%$,\newline  $(\alpha=1.14,\beta=0.08)$\\\hline

 \end{tabular}
  \caption{Comparison of quality factors obtained for different scalings.}
  \label{tab:Muppidiscalings}
  \end{center}
\end{table}

Table \ref{tab:Muppidiscalings} summarizes the different scalings and how successfully they collapse the data.
 The proposed scaling achieves similar or better collapse than the scaling proposed by \cite{Muppidi2005}. It requires the determination of two parameters, whereas the $rd (\frac{h}{d})^C$ scaling  requires only one. On the other hand $h$ has to be extracted from the data independently for each trajectory, whereas $r_m$ and $\delta$ are experimental parameters known \textit{a priori}.\hspace{1mm} For this data, the scaling suggested in \cite{Gutmark2008} does not result in trajectory collapse,  on the contrary it increases the dispersion of the curves. It is most likely due to a typographic mistake in the printed scaling formula. For instance, the $\frac{d}{\delta}$ factor has to be inverted to make physical sense.

\subsection{Scaling of CVP trajectories}
\label{sec:17}

Experimental CVP trajectory data analyzed in section \ref{sec:9} and jet trajectory data discussed in section \ref{sec:16} show that both types of trajectories behave in the  same manner when parameters vary. This is to be expected since the CVP is a structure created by the jet and it seems CVP trajectory follows jet centerline trajectory.\\

Nevertheless there are differences between CVP and jet trajectories. CVP trajectories do not start at the jet exit ($x=0,y=0$) and are lower than jet trajectories. Moreover, since jet trajectories (\cite{Salewski2007}) and CVP trajectories are parallel in the far field, it is impossible for both types of trajectories to assume a power law and retain that parallelism. Nevertheless a power law will be used to scale CVP trajectories, keeping in mind that the starting point abscissa for CVP trajectories vary ($x_{start} \simeq 1.5d$ for most trajectories). \\

Since the trajectories of the CVP are influenced by momentum ratio $r_m$, diameter $d$ and boundary layer thickness $\delta$,  the scaling described in equation \ref{eq:Newscl} is applied to our data.

\subsection{Determination of the optimal scaling for CVP trajectories}
\label{sec:18}
To determine the influence of boundary layer thickness on CVP trajectories we consider configurations 9 to 13. In these cases, boundary layer thickness varies from $1.36d$ to $2.26d$ while $r_m=2.2 \pm 5\%$. Best collapse is obtained for $\beta=0.91$, thus for these cases boundary layer thickness has a significant relative influence on CVP trajectory. \\

\begin{figure}
\centering
\includegraphics[width=0.65\textwidth]{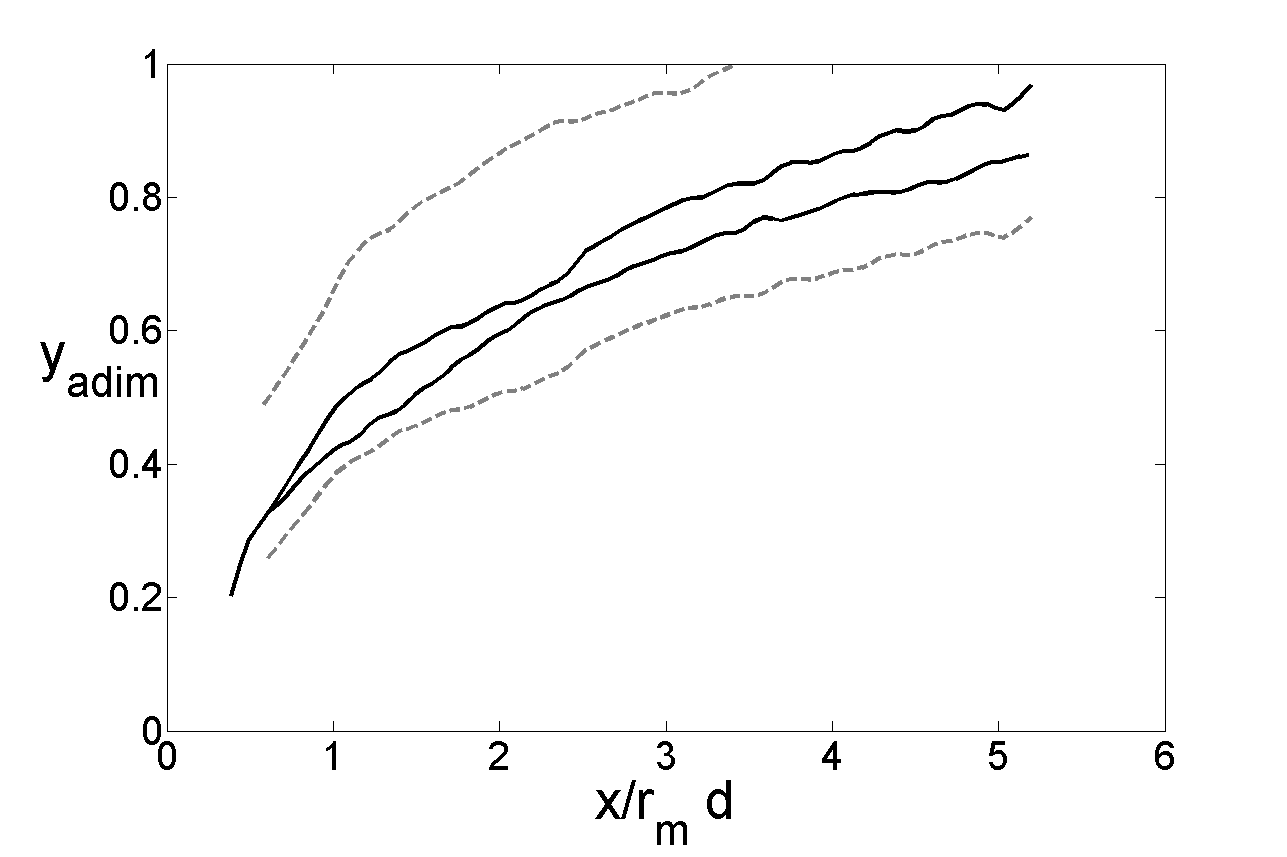}
\caption{ Highest and lowest trajectories for configurations 9 to 13, before (grey, dotted line) and after (black, solide line) scaling by $d(\delta/d)^{\beta}$ ($Q=28.3\%$). $y$ axis is normalized to help comparison.}
\label{fig:11to15delta}
\end{figure}

Fig. \ref{fig:11to15delta} shows the highest and lowest of the trajectories before and after scaling by $d (\frac{\delta}{d})^{\beta}$. Collapse is significant ($Q=28.3\%$).  $\beta=0.91$ is much higher than the value found in section \ref{sec:11} although the velocity ratios are comparable ($r=1.5$ and $r=1.6$).  CVP trajectories being lower, interaction with boundary layer would be stronger for this velocity ratio, resulting in a higher value for $\beta$.

To go further, a simplifying assumption is made: $\beta$ is assumed to be constant. In other words, the way the boundary layer thickness affects trajectory is considered  independent of other parameters such as $r$, $r_{m,jet}$ or $r_{m,cf}$. 
Of course, this is not strictly true as shown in section \ref{sec:16}.   Furthermore since some of these trajectories are close or even inside the boundary layer (see Fig. \ref{fig:trajsr}) it stands to reason $\beta$ would change with momentum ratio.  However data are insufficient for a more thorough analysis of this issue,   another extensive parametric study would be required. Nevertheless, based on the data from \cite{Muppidi2005} analyzed in the previous section, we can expect $\beta$ to be a decreasing function of $r_m$.

\begin{figure}
\centering
\includegraphics[width=0.65\textwidth]{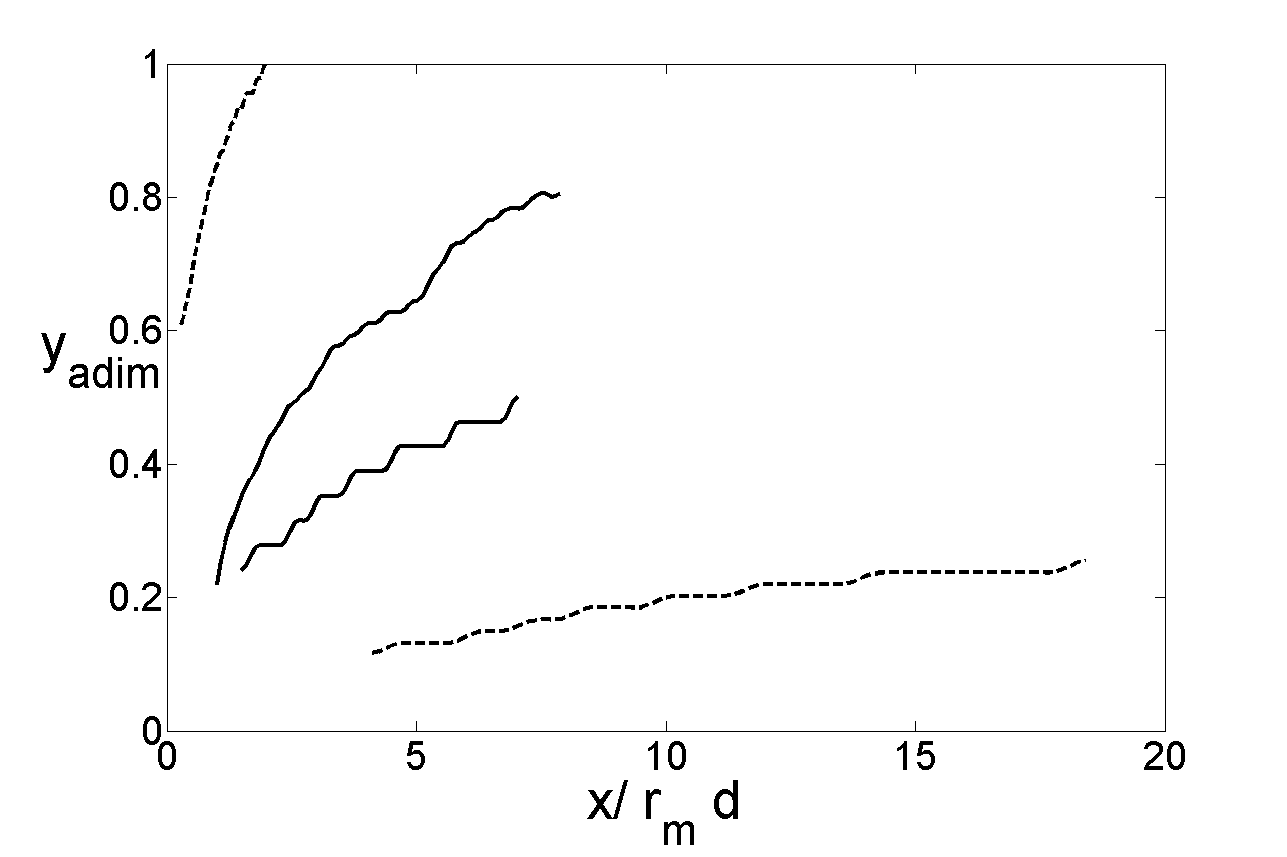}
\caption{ Lowest and highest CVP trajectories for configurations 1 to 8 and 14 to 22. No scaling (- -) and after scaling (-) leading to $Q=13.1\%$. $y$ axis is normalized to allow visual comparison.}
\label{fig:1to1016to24R}
\end{figure}

To determine the influence of $r_m$ we consider configurations 1 to 8 and 14 to 22. All these configurations feature variations in $r_m$. However these variations are brought about in different ways: variations in  $r$ $(0.51<r<3.05)$ by changing jet velocity and cross flow velocity and variations in $r_{m,jet}$ by changing jet velocity profile. We find $\alpha=1.23$ ($Q=13.1\%$). Fig. \ref{fig:1to1016to24R} shows the highest and lowest trajectories before and after scaling by $r_m^{\alpha} d (\frac{\delta}{d})^{\beta}$.

For all 22 configurations the scaling mentioned gives $Q=13.14\%$. This relatively high value is most likely due to the fact that $\alpha$ and $\beta$ are considered constant when they have been shown to depend on $r$. Furthermore the underlying assumption of a power law scaling, i.e that there exists a scaling  such that trajectories can be expressed as $y=A x^b$ where $A$ and $b$ are constant, is erroneous as shown by \cite{New2006}. However the proposed scaling does allow for significant collapse with a $43\%$ improvement over the $rd$ scaling.

Using the computed values for $\alpha$ and $\beta$, the scaled data are well fitted with a power law as described in equation \ref{eq:trjct}:
\begin{equation}
\frac{y}{{r_m}^{\alpha} d (\frac{\delta}{d})^{\beta}}=A(\frac{x}{r_m d})^b
\label{eq:trjct}
\end{equation}

  with $A=0.48$, $b=0.42$.  Commonly $rd$ scalings of the jet trajectory yield $1.2<A<2.6$ and $0.28<b<0.34$.  For CVP trajectories $A$ is lower because the trajectory lies under the jet centerline. Possible uses of this equation are many-fold. For example when devising an experiment involving jets in cross-flow it could be helpful to choose the proper geometrical and physical parameters for a given objective.

\begin{figure}
\centering
\begin{tabular}{c c}
\includegraphics[width=0.5\textwidth]{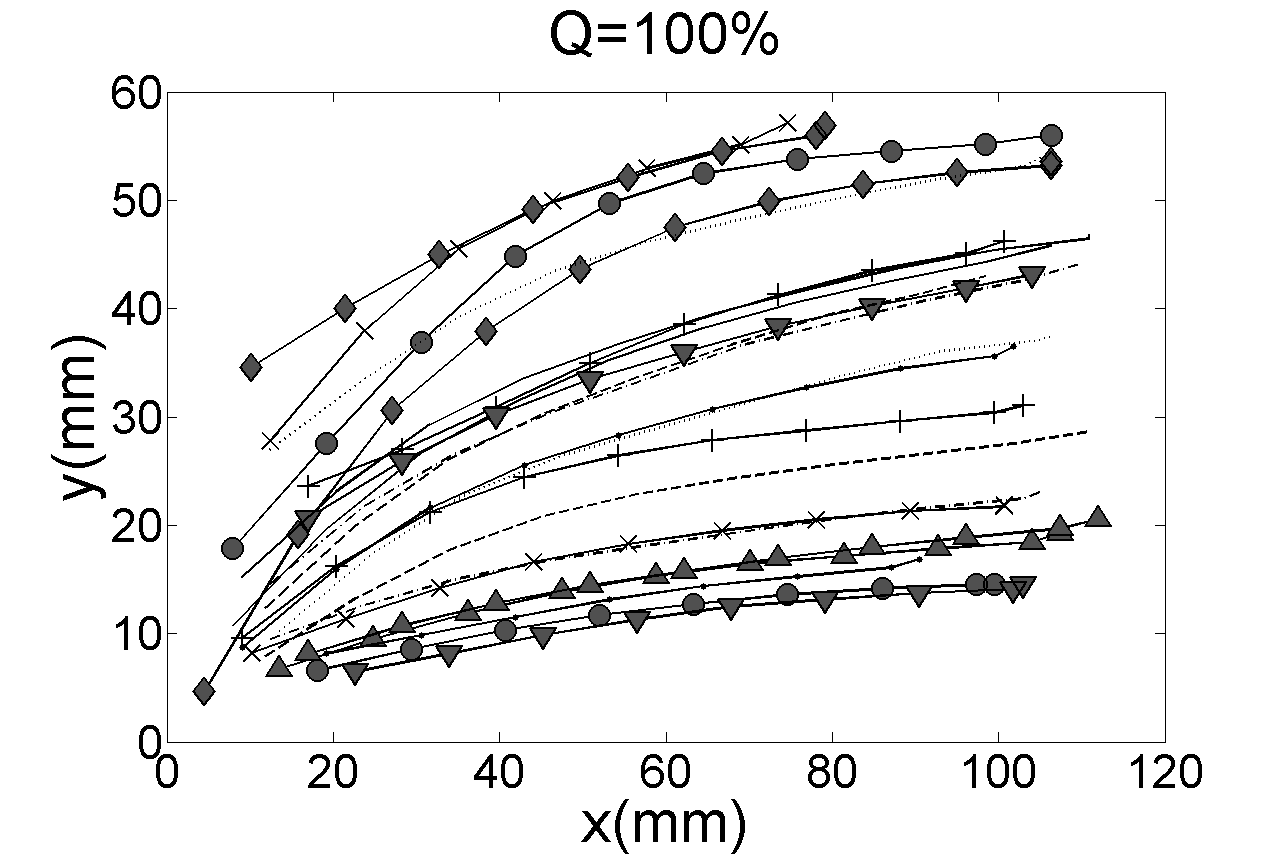} & \includegraphics[width=0.5\textwidth]{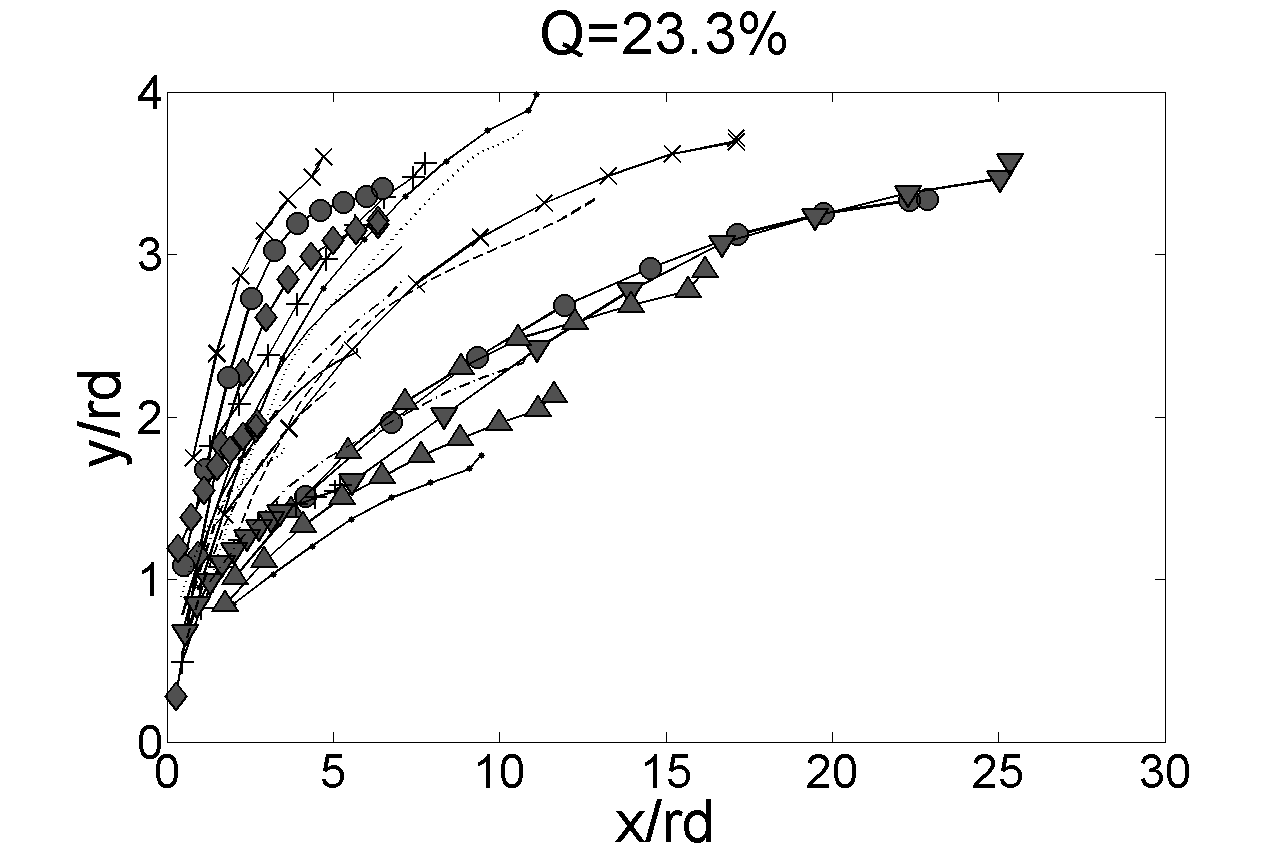}\\
a) & b) \\
\includegraphics[width=0.5\textwidth]{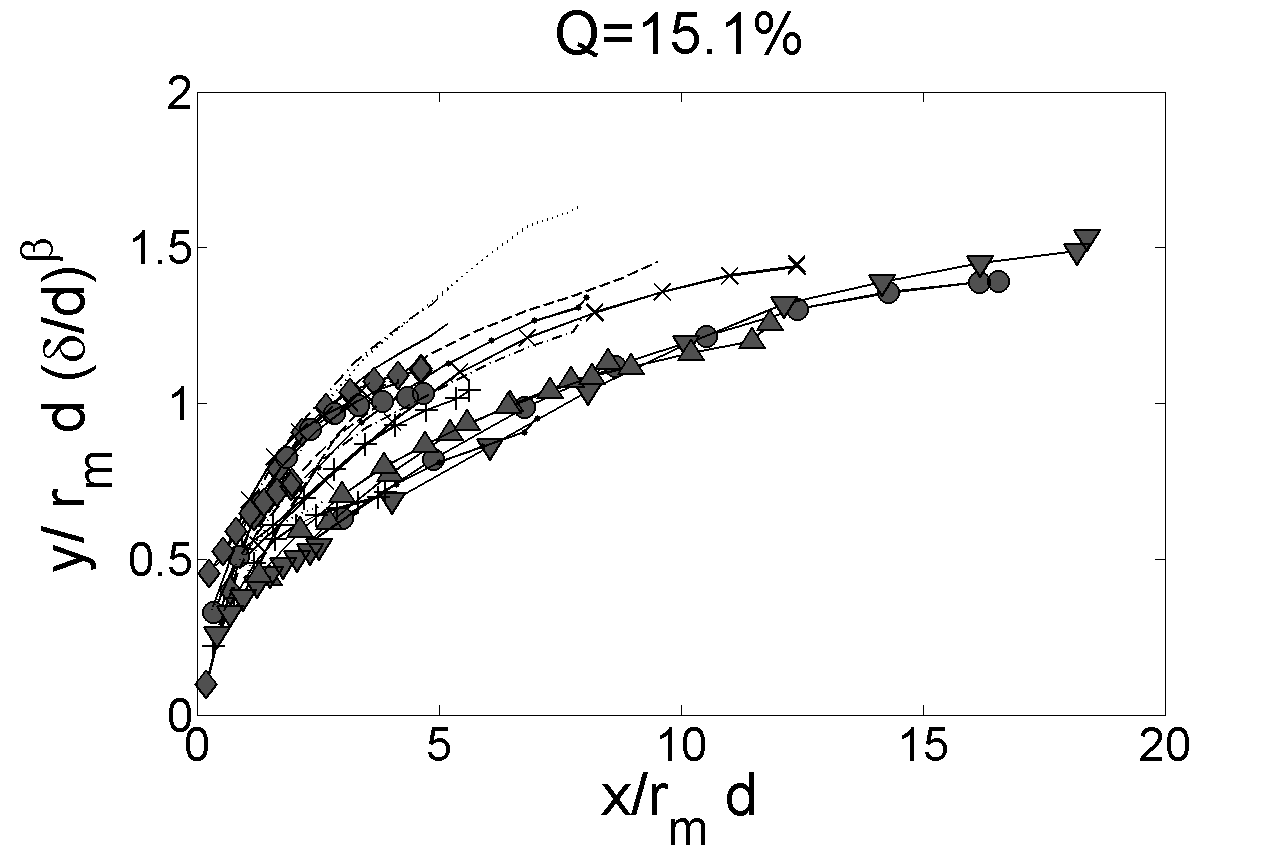} &\includegraphics[width=0.5\textwidth]{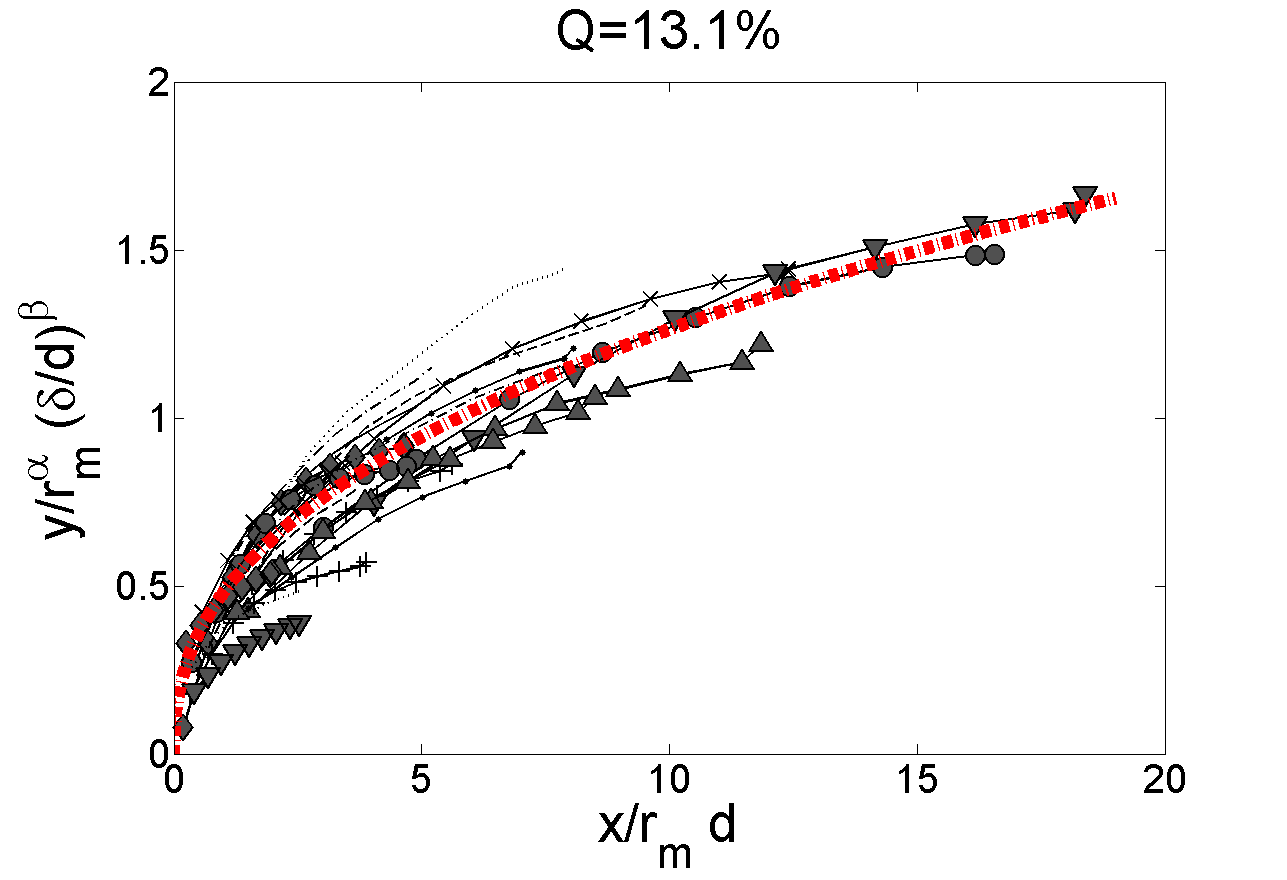}\\ 
c) & d)

\end{tabular}
\caption{Whole set of trajectories scaled with different scalings. $y$ axis is normalized to help comparison. a) Trajectories without scaling, b) Trajectories with $rd$ scaling, c) Trajectories with $(\delta/d)^{\beta}$ scaling, d) final scaling together with the trajectory described in equation \ref{eq:trjct}  (red dotted line).}

\label{fig:allenvelopes}
\end{figure}

Finally, we summarize on Fig. \ref{fig:allenvelopes} how the main scalings discussed previously collapse all CVP trajectories. The improvements in collapse brought about by each scalings are clear and quantified by the decrease of the quality factor which is minimum for the scaling based on momentum ratio proposed in this study (Fig. \ref{fig:allenvelopes}d).  For the range of parameters considered here, equation \ref{eq:trjct} allows for a decent approximation of the CVP's position in space as illustrated on Fig. \ref{fig:allenvelopes} d.\\

\section{Conclusions}
\label{sec:19}
An experimental study of the CVP trajectories created by a round JICF has been carried out in an hydrodynamic tunnel. 3D3C velocity fields were used to identify the CVP's and their corresponding outflow regions. The outflow region is used to define and compute CVP trajectories for 22 JICF configurations, including those with a low velocity ratio $r$. The influence of jet velocity  and profile as well as cross flow velocity and boundary layer thickness on CVP trajectories is investigated. Parallels are drawn between the behavior of jet and CVP trajectories.\\

 A more general momentum ratio $r_m$  is introduced as an improvement of the velocity ratio $r$ to take into account the boundary layer and jet exit momentum distributions. The relevance of $r_m$ for jet and CVP trajectories is demonstrated for numerical and experimental data.
\\
Experimental CVP trajectories and jet trajectories from the literature are scaled and analyzed. The quality of a given scaling is defined and allows for the determination of the relative significance of each parameter (momentum ratio, boundary layer thickness) on trajectories. A new scaling taking into account  jet exit momentum distributions, velocity ratio and  boundary layer thickness is proposed.  
 \\
 Finally, a unique trajectory taking into account all relevant parameters is suggested for CVP trajectories. 

\section{Acknowledgments}

The authors gratefully acknowledge the ADEME (Agence De l'Environnement et la Maitrise de l'Energie) for its financial support, as well the reviewers for their helpful comments.

\bibliographystyle{unsrt}	
\bibliography{LowVelocityRatioScaling}

\end{document}